\documentclass{PoS}

\usepackage{graphicx}
\usepackage{amsmath}
\usepackage{subfigure}
\usepackage{wrapfig}
\usepackage{epsfig}

\newcommand{\be}{\begin{equation}}
\newcommand{\ee}{\end{equation}}
\newcommand{\bea}{\begin{eqnarray}}
\newcommand{\eea}{\end{eqnarray}}
\newcommand{\nn}{\nonumber \\}

\def\Dslash{D\hskip-0.65em /}

\title{New Higgs physics from the lattice\thanks{Based on contributions from K.~Holland, J.~Kuti, D.~Nogradi, and C.~Schroeder.}}

\ShortTitle{New Higgs physics from the lattice}

\author{Zoltan Fodor\\
 Department of Physics, University of Wuppertal, Gauss Strasse 20, D-42119, Germany\\
 Email: \email{fodor@bodri.elte.hu}}

\author{Kieran Holland\\
Department of Physics, University of the Pacific\\
3601 Pacific Ave, Stockton CA 95211, USA\\
Email: \email{kholland@pacific.edu}}

\author{Julius Kuti \\
        Department of Physics 0319, University of California, San Diego\\
        9500 Gilman Drive, La Jolla, CA 92093, USA\\
        E-mail: \email{jkuti@ucsd.edu}}

\author{Daniel Nogradi\\
 Department of Physics, University of Wuppertal, Gauss Strasse 20, D-42119, Germany\\
 Email: \email{nogradi@lorentz.leidenuniv.nl}}

\author{Chris Schroeder\\
        Department of Physics 0319, University of California, San Diego\\
        9500 Gilman Drive, La Jolla, CA 92093, USA\\
        E-mail: \email{crs@physics.ucsd.edu}}

\abstract{
We report the first results from our comprehensive lattice tool set
to explore non-perturbative aspects of Higgs physics 
in the Standard Model.
We demonstrate in Higgs-Yukawa models that Higgs mass lower bounds and 
upper bounds can be determined in lattice simulations when 
triviality requires the necessity of a finite cutoff to maintain non-zero
interactions. The vacuum instability problem is investigated and the
lattice approach is compared with the traditional
renormalization group procedure which sets similar goals to correlate 
lower and upper Higgs mass bounds with the scale of new physics. 
A novel feature of our lattice simulations is the use 
of Ginsparg-Wilson fermions to represent the effects of Top quark loops in 
Higgs dynamics. The need for chiral lattice fermions is discussed and the
approach is extended to full Top-Higgs-QCD dynamics. 
We also report results from our large $N_F$ analysis of Top-Higgs Yukawa models to
gain analytic insight and to verify our new lattice tool set which is deployed
in the simulations.
The role of non-perturbative lattice studies to investigate heavy Higgs
particle scenarios is illustrated in extensions of the Standard Model.
}

\FullConference{The XXV International Symposium on Lattice Field Theory\\
		 July 30 - August 4 2007\\
		 Regensburg, Germany}

\begin{document}

\section{Introduction}

The search for the Higgs has become a major issue in particle
physics as the LHC is nearing its completion. The Standard Model (SM) cannot be
considered complete given that the Higgs is as-yet unobserved and it
is not clear how Electroweak symmetry is broken in nature. If the
Higgs is seen, its properties could tell us about physics beyond the
Standard Model, such as the energy scale of a more fundamental
theory. The current lower bound for the Higgs mass from direct
searches is $114.4$~GeV~\cite{Barate:2003sz}. The Higgs mass can also
be inferred indirectly by fitting the Standard Model to a host of
Electroweak precision measurements. The best perturbative fit gives $m_H=76^{+33}_{-24}$~GeV, 
so the data certainly seem to prefer the Higgs to be light~\cite{LEP2005}. 
However, the global fitting procedure, which favors a surprisingly
low Higgs mass, has its own intrinsic issues, 
perhaps a hint that deviations from the Standard
Model are already present~\cite{Chanowitz:2002cd}. 
Larger Higgs masses together with new physics threshold effects
at the TeV scale will require new extended 
analysis~\cite{Peskin:2001rw,Grojean:2006nn} where non-perturbative effects
may come into play.

Based on the assumption that the Standard Model is only valid up 
to some energy scale,
lower and upper bounds on the Higgs mass were established before 
without relying on input from Electroweak precision measurements.
Bounds on the Higgs mass
are valuable for two reasons. Firstly, they cut down the parameter
space where one searches for a Standard Model Higgs. Secondly, if the
Higgs is found, measuring its mass and knowing the bounds it must obey
would indicate the maximum energy scale up to which the Standard Model
can work.
In phenomenology, the origin of the
lower bound is thought to be the vacuum instability the Top quark loop would generate, if 
the Higgs mass were too light. The upper bound in phenomenological analysis
is simply calculated by not
allowing the running Higgs coupling $\lambda(t)$ to become strong at the
cutoff scale $\Lambda$ which represents new physics before $\lambda(t)$ 
would run into the fictitious Landau pole. 
These ideas on lower and upper Higgs mass bounds have been applied to 
the Standard Model for almost 30 years and have
been increasingly refined. 

\FIGURE[t]{
\includegraphics[width=6cm,angle=90]{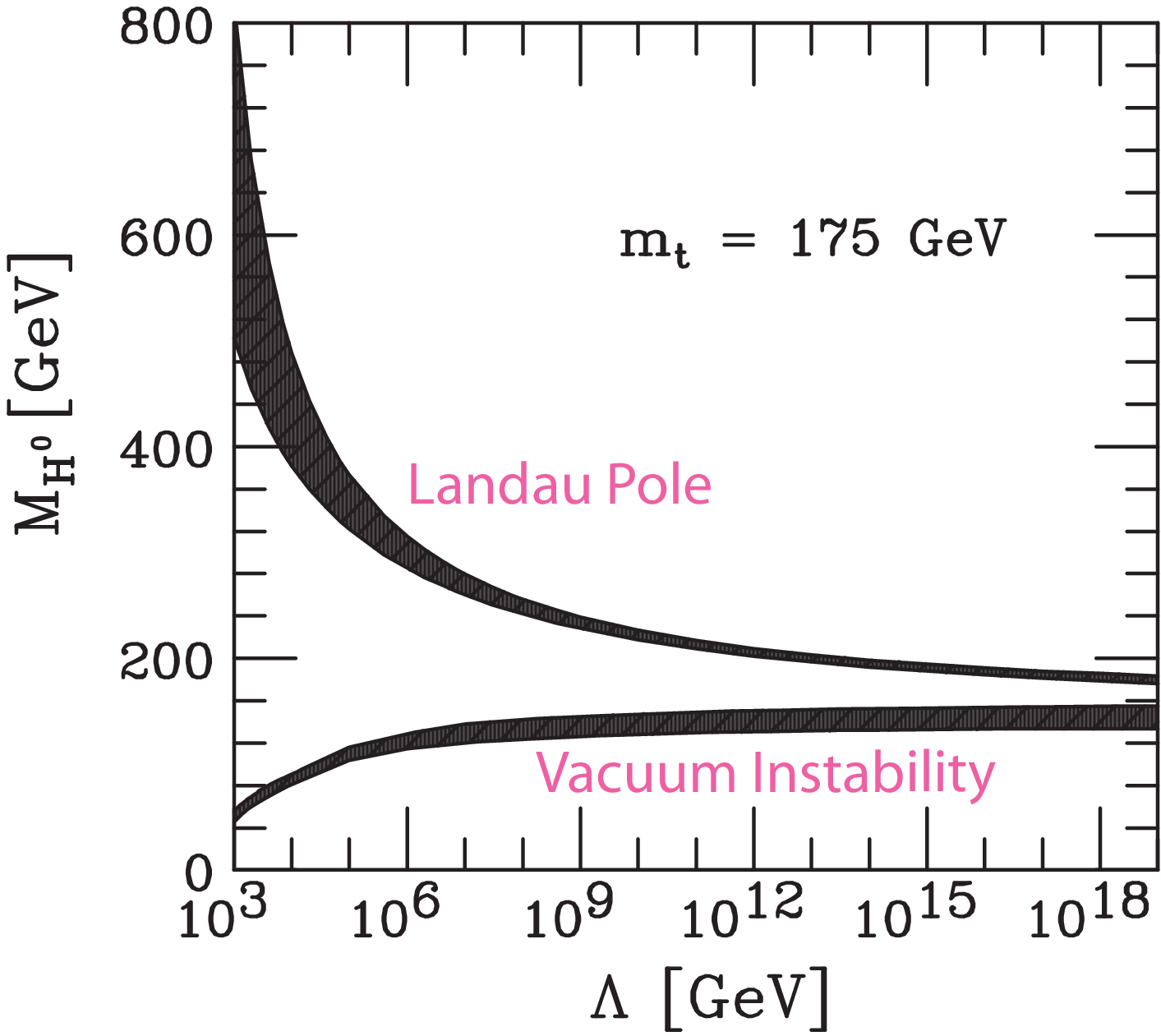}
\caption{\label{fig:PDG}
Upper and lower bounds for the Higgs mass as a function of the scale
of new physics beyond the Standard Model, from \cite{Hagiwara:2002fs}.}
}
The bounds given by the state-of-the-art
calculations were reviewed in~\cite{Hagiwara:2002fs} 
and shown in Figure~\ref{fig:PDG},
based on the original work in  \cite{Hambye:1996wb} and
\cite{Casas:1996aq}. There are several things one can learn from this
plot. The Standard Model apparently cannot generate a Higgs boson
heavier than 1~TeV without strong Higgs self-interactions and a low threshold for
new physics in the TeV range,
a scenario not consistent with the perturbative loop expansion of the
Electroweak precision analysis. What non-perturbative 
modifications on the TeV scale would
support a heavy Higgs particle, consistent with Electroweak precision data, 
is one of the motivations for our lattice studies~\cite{Peskin:2001rw,Grojean:2006nn}.
The lower bound is interesting for today's phenomenology, given
the current experimental limits. If the Higgs mass is around 100~GeV, this would
intersect with the lower bound in Figure~\ref{fig:PDG} somewhere between 10 and 100~TeV,
beyond which apparently new physics should enter. 

One major goal of our lattice Higgs project is to understand the role of vacuum instability
and the Landau pole in an exact non-perturbative setting when the intrinsic cutoff 
in the Higgs sector is not removable and low in the TeV range.
Another goal is to explore the role of non-perturbative
Higgs physics from the lattice in extensions of the perturbative SM analysis, including the possibility of a heavy Higgs
particle within the Higgs reach of the LHC.

The outline of this paper is as follows.
In section 2 we will report results from the large $N_F$ analysis of the Top-Higgs Yukawa
model of a single real scalar field coupled to $N_F$ fermions. The 
influence of the non-removable intrinsic
cutoff (triviality) on the exact renormalization group (RG) flow is exhibited. 
The vacuum instability problem of the model is discussed on the lattice in section 3
and compared with the traditional
renormalization group procedure of the Standard Model
(earlier versions of this work on vacuum instability have been discussed in
\cite{Holland:2003jr} and \cite{Holland:2004sd}). 
In section 4 we present the Wilsonian view on the renormalization group as applied to the 
vacuum instability and Higgs lower bound problems.
The first lattice simulation results on the 
Higgs mass lower bound, using chiral
lattice fermions in Top-Higgs Yukawa models, are reported in section 5.

Using the higher derivative (Lee-Wick) extension 
of the Higgs sector~\cite{Jansen:1993jj,Jansen:1993ji,Grinstein:2007mp},
we will illustrate in section 6 how non-perturbative
lattice studies might help to investigate heavy Higgs particle
scenarios in the 500-800 GeV Higgs mass range relevant for future LHC physics. 
Constraints from Electroweak precision data on the heavy Higgs particle are briefly discussed.

\section{Top-Higgs Yukawa model in large $\rm\bf{N_F}$ limit}
For pedagogical purposes, we first consider a Higgs-Yukawa model of a
single real scalar field coupled to $N_F$ massless fermions. The saddle point approximation
in the large $N_F$ limit becomes exact and
this will allow us to demonstrate that the theory is
trivial. We will also calculate the flow of the renormalized
couplings as a function of the energy scale to identify
problems with the vacuum instability scenario when the intrinsic
cutoff is non-removable. 
Similar behavior is expected at finite $N_F$ which requires non-perturbative 
lattice simulations.

\subsection{Renormalization scheme}

Let us start with the bare Lagrangian of the Higgs-Yukawa theory in
Euclidean space-time, which is 
\be
{\cal L} = \frac{1}{2} m_0^2 \phi_0^2 + \frac{1}{24} \lambda_0
\phi_0^4 + \frac{1}{2}\left(\partial_\mu \phi_0\right)^2
+ \bar{\psi}^a_0 \left(\gamma_\mu \partial_\mu + y_0
\phi_0\right) \psi^a_0, 
\label{eq:bareL}
\ee
where $a=1,...,N_F$ sums over the degenerate fermion flavors and the
subscript $0$ denotes bare quantities. We rewrite this as
\bea
{\cal L} &=& \frac{1}{2} m_0^2 Z_\phi \phi^2 + \frac{1}{24} \lambda_0
Z_\phi^2 \phi^4 + \frac{1}{2} Z_\phi \left(\partial_\mu \phi\right)^2
+ Z_\psi \bar{\psi}^a \left(\gamma_\mu \partial_\mu + y_0
\sqrt{Z_\phi} \phi\right) \psi^a \nn
&=& \frac{1}{2} (m^2 + \delta m^2) \phi^2 + \frac{1}{24} (\lambda
+ \delta \lambda) \phi^4 + \frac{1}{2} (1 + \delta z_\phi)
\left(\partial_\mu \phi\right)^2 \nn
&& + (1 + \delta z_\psi)
\bar{\psi}^a \gamma_\mu \partial_\mu \psi^a + \bar{\psi}^a (
y + \delta y) \phi \psi^a,
\label{eq:renormL}
\eea
where we have introduced the wavefunction renormalization factors
$
Z_\phi = 1 + \delta z_\phi, ~ Z_\psi = 1 + \delta z_\psi
$
and renormalized parameters with their corresponding counterterms. The
connections between the bare and renormalized parameters are
\be
m_0^2 Z_\phi = m^2 + \delta m^2, \hspace{0.5cm} \lambda_0 Z_\phi^2 =
\lambda + \delta \lambda, \hspace{0.5cm} Z_\psi \sqrt{Z_\phi} y_0 = y
+ \delta y.
\label{eq:barerenorm}
\ee
In the limit where $N_F$
becomes large, all Feynman diagrams with Higgs loops are suppressed
relative to those with fermion loops. Hence, two of the counterterms
vanish, 
$
\delta y = 0, ~ \delta z_\psi = 0,
$
as there are no radiative corrections to the fermion propagator or to
the Higgs-fermion coupling. Let us specify the renormalization
conditions which determine the remaining counterterms.

In the large $N_F$ limit, the renormalized Coleman-Weinberg effective
potential \cite{Coleman:1973jx} is 
\be
U_{\rm eff} = \frac{1}{2} m^2 \phi^2 + \frac{1}{24} \lambda \phi^4 +
\frac{1}{2} \delta m^2 \phi^2 + \frac{1}{24} \delta \lambda \phi^4 - 2
N_F \int_k \ln[ 1 + y^2 \phi^2/k^2]
\label{eq:Ueff}
\ee
containing the tree-level contributions from the renormalized
parameters and their counterterms, and the infinite sum of all
diagrams with one fermion loop and an even number of external $\phi$
legs. The factor $N_F$ comes from all the possible fermions
which can appear in the single loop and we use the notation 
$\frac{1}{(2\pi)^4}\int d^4k \rightarrow$ $\int_k$ for loop integrals.
The vacuum expectation value
$\phi=v$ is where $U_{\rm eff}$ has an absolute minimum i.e.~$U'_{\rm
  eff}(v)=0$. In the Higgs phase of the theory, $v \ne 0$. At
tree-level, this gives the relation  
\be
m^2 + \frac{1}{6} \lambda v^2 = 0,
\label{eq:tree}
\ee
coming from the first two terms in Equation (\ref{eq:Ueff}). Our first
renormalization condition is that we want to maintain the tree-level
relation in Equation (\ref{eq:tree}) exactly, giving
\be
\delta m^2 + \frac{1}{6} \delta \lambda v^2 - 4 N_F y^2 \int_k
\frac{1}{k^2 + y^2 v^2} = 0.
\label{eq:renorm1}
\ee
The counterterms exactly cancel all the finite and infinite
contributions of the radiative diagrams. The same relation can also be
determined by demanding that the tadpole diagram is exactly cancelled by
the counterterms.

In the Higgs phase, we define the Higgs fluctuation around the vev as
$\phi = \varphi + v$. At tree-level, the mass of the Higgs fluctuation
i.e.~$U''_{\rm eff}(v)$ is
\be
m_H^2 = m^2 + \frac{1}{2} \lambda v^2 = \frac{1}{3} \lambda v^2.
\label{eq:mhiggs}
\ee
In the large $N_F$ limit, the inverse propagator of the Higgs
fluctuation is
\bea
G^{-1}_{\varphi \varphi}(p^2) &=& p^2 + m^2 + \frac{1}{2} \lambda v^2 +
p^2 \delta z_\phi + \delta m^2 + \frac{1}{2} \delta \lambda v^2 -
\Sigma(p^2) \nn
\Sigma(p^2) &=& - 4 N_F y^2 \int_k \frac{y^2 v^2 - k.(k-p)}{(k^2 + y^2
  v^2)((k-p)^2 + y^2 v^2)},
\label{eq:Gphiphi}
\eea
where all Higgs-loop diagrams are suppressed relative to the single
fermion-loop diagram. We impose the condition that
\be
G^{-1}_{\varphi \varphi}(p^2 \rightarrow 0) = p^2 + m_H^2,
\label{eq:renormG}
\ee
which separates into two renormalization conditions:
\be
\delta m^2 + \frac{1}{2} \delta \lambda v^2 - \Sigma(p^2=0) = 0
\label{eq:renorm2}
\ee
and
\be
\delta z_\phi - \left. \frac{d \Sigma(p^2)}{d p^2} \right|_{p^2=0} =
0.
\label{eq:renorm3}
\ee
The renormalization condition Equation (\ref{eq:renorm2}) maintains
the tree-level relation in Equation (\ref{eq:mhiggs}) exactly. Again,
the counterterms precisely cancel all the finite and infinite radiative
contributions. We should point out that the Higgs mass defined as the
zero-momentum piece of $G^{-1}_{\varphi \varphi}$ is identical to
that defined via the curvature $U''_{\rm eff}(v)$. This is not the
same as the true physical mass given by the pole of the propagator,
and these masses can be related to one another in perturbation theory.

The renormalization conditions Equations (\ref{eq:renorm1}) and
(\ref{eq:renorm2}) can easily be solved.
Because we wish to demonstrate triviality in this
theory, we use some finite cutoff in the momentum integrals and
examine what occurs as this cutoff is removed. 
We will use a simple hard-momentum cutoff $|k| \le
\Lambda$. Exactly the same conclusions would be reached using instead
e.g.~Pauli-Villars regularization. The non-zero counterterms after
the loop integration are 
\bea
\delta m^2 &=& \frac{N_F y^2}{2 \pi^2} \left[ \frac{1}{2} \Lambda^2 +
  \frac{y^4 v^4}{2(\Lambda^2 + y^2 v^2)} - \frac{1}{2} y^2 v^2 \right] ~,
\nn
\delta \lambda &=& - \frac{3 N_F y^4}{\pi^2} \left[ \frac{y^2
    v^2}{2(\Lambda^2 + y^2 v^2)} - \frac{1}{2} - \frac{1}{2} \ln\left(
    \frac{y^2 v^2}{\Lambda^2 + y^2 v^2} \right) \right] ~, \nn
\delta z_\phi &=& - \frac{N_F y^2}{2 \pi^2} \left[ \frac{1}{4} \ln
  \left( \frac{y^2 v^2 + \Lambda^2}{y^2 v^2} \right) + \frac{-5
    \Lambda^4 - 3 \Lambda^2 y^2 v^2}{12(\Lambda^2 + y^2 v^2)^2}
\right].
\label{eq:solve2}
\eea
As we said earlier, in the large $N_F$ limit, the fermion inverse
propagator receives no radiative correction,
\be
G^{-1}_{\psi \psi}(p) = p_\mu \gamma_\mu + yv,
\label{eq:Gferm}
\ee
so we identify the fermion mass as $m_t=yv$ (looking ahead to the Top
quark), which we substitute into all of the above equations. 

\subsection{Triviality}
Let us first consider the regime $m_t/\Lambda \ll 1$, where the cutoff
is much larger than the physical scale. In this limit, we get
\be
Z_\phi = \left[ 1 + \frac{N_F y_0^2}{8 \pi^2} \left( \ln
    \left[\frac{\Lambda^2}{m_t^2}\right] - \frac{5}{3} \right)
\right]^{-1}.
\label{eq:Zvanish}
\ee
For any finite bare Yukawa
coupling $y_0$, the Higgs wavefunction renormalization factor $Z_\phi$
vanishes logarithmically as the cutoff is removed, $m_t/\Lambda
\rightarrow 0$. 
This same logarithmic behavior, for any choice of bare couplings, will appear in all of
the renormalized couplings, leading to the triviality scenario:
a finite cutoff must be kept to maintain
non-zero interactions. Explicitly, the renormalized Yukawa coupling is
\be
y^2 = y_0^2 Z_\phi = y_0^2 \left[ 1 + \frac{N_F y_0^2}{8 \pi^2} \left( \ln
    \left[\frac{\Lambda^2}{m_t^2}\right] - \frac{5}{3} \right)
\right]^{-1} 
\rightarrow \left[\frac{N_F}{8 \pi^2} \ln
  \frac{\Lambda^2}{m_t^2}\right]^{-1}, \hspace{0.5cm} {\rm
  as}~\frac{m_t}{\Lambda} \rightarrow 0.
\label{eq:yvanish}
\ee
%
%
For the renormalized Higgs coupling, we have
\bea
\lambda &=& \lambda_0 Z_\phi^2 - \delta \lambda = \lambda_0 Z_\phi^2 +
\frac{3 N_F y^4}{\pi^2} \left[ \frac{m_t^2}{2(\Lambda^2 + m_t^2)} -
  \frac{1}{2} - \frac{1}{2} \ln \left(\frac{m_t^2}{\Lambda^2 +
      m_t^2}\right) \right] \nn
&\rightarrow& Z_\phi^2 \left[ \lambda_0 + \frac{3 N_F y_0^4}{\pi^2}
  \left( - \frac{1}{2} - \frac{1}{2} \ln \frac{m_t^2}{\Lambda^2}
  \right) \right] 
\rightarrow 12 \left[\frac{N_F}{8 \pi^2} \ln
  \frac{\Lambda^2}{m_t^2}\right]^{-1}, \hspace{0.5cm} {\rm
  as}~\frac{m_t}{\Lambda} \rightarrow 0.
\label{eq:lamvanish}
\eea
The slow logarithmic vanishing of $y$ and $\lambda$ allows to
have a relatively large separation of cutoff and physical scales and
still maintain significant interactions. 
However, the standard renormalization procedure of
removing the cutoff completely gives a non-interacting theory.
Although completely unphysical, we can also consider the limit
$m_t/\Lambda \gg 1$, where the cutoff is much below the physical
scale. From Equation (\ref{eq:solve2}), we see this gives
$
\delta \lambda = 0, ~ \delta z_\phi = 0,
\label{eq:counterzero2}
$
and hence $Z_\phi \rightarrow 1$. In this limit, the connection
between bare and renormalized parameters is simply
$
\lambda = \lambda_0, ~ y = y_0.
\label{eq:barerenorm2}
$
This result is not surprising: deep in the cutoff regime, we simply
have the bare theory, with no separation into renormalized parameters
and their counterterms. This will be relevant when we discuss whether
the vacuum can become unstable. 

\subsection{Renormalization group flow}
The physical properties of the theory are fixed as soon as one chooses
a complete set of bare parameters. As the cutoff is varied, the
renormalized couplings flow in order to maintain exactly the
renormalization conditions we have imposed. Using the explicit cutoff
dependence of $y$ and $\lambda$, we can calculate this Callan-Symanzik
flow. In the limit $m_t/\Lambda \ll 1$, from
Equations (\ref{eq:yvanish}) and (\ref{eq:lamvanish}), we have
\bea
\Lambda \frac{d y^2}{d \Lambda} &=& - y_0^2 Z_\phi^{2} \frac{N_F
  y_0^2}{4 \pi^2} = - \frac{N_F y^4}{4 \pi^2}~, \nn
\Lambda \frac{d \lambda}{d \Lambda} &=& \frac{1}{16 \pi^2} \left[ - 8
  N_F \lambda y^2 + 48 N_F y^4 \right]~.
\label{eq:contflow}
\eea
The same $\beta$ functions would be obtained in the
large $N_F$ limit for the running $y$ and $\lambda$
couplings in scale dependent RG flows using e.g.~dimensional regularization, 
where no cutoff would explicitly appear. (Since increasing 
$\Lambda$ corresponds to decreasing mass scale $\mu$, the 
$\beta$ functions in Equation~(\ref{eq:contflow}) have opposite signs).
It is important to note that the two
RG schemes have very different physical meanings: Equation~(\ref{eq:contflow})
describes the response to changing the cutoff whereas the scale dependent RG flow
compensates for the arbitrary choice of the renormalization scale
at finite cutoff.
When the cutoff is far above the physical scales, the
finite cutoff effects are negligible and we expect to reproduce the unique
cutoff-independent $\beta$ functions. However, as the cutoff is
reduced and $m_t/\Lambda$ increases, this cannot continue to hold
indefinitely, as the renormalized couplings must eventually flow to
the bare ones, as explained above.

Let us demonstrate an explicit example of the Callan-Symanzik RG
flow in the presence of a finite cutoff. 
In the large $N_F$ limit, $m_t = yv = y_0 v_0$. The bare vev is
determined by the minimum of the bare effective potential
\be
U_{\rm eff, 0} = \frac{1}{2} m_0^2 \phi_0^2 + \frac{1}{24} \lambda_0
\phi_0^4 - 2 N_F \int_k \ln\left[ 1 + y_0^2 \phi_0^2/k^2 \right].
\label{eq:bareUeff}
\ee
Using a hard-momentum cutoff, this gives
\be
m_0^2 + \frac{1}{6} \lambda_0 v_0^2 - \frac{N_F y_0^2}{2 \pi^2} \left[
  \frac{1}{2} \Lambda^2 + \frac{1}{2} y_0^2 v_0^2 \ln \left(\frac{y_0^2
    v_0^2}{\Lambda^2 + y_0^2 v_0^2}\right) \right] = 0.
\label{eq:gap}
\ee
We express all dimensionful quantities in units of the cutoff
$\Lambda$. We pick some fixed values for $\lambda_0$ and
$y_0$. Varying the value of $m_0^2/\Lambda^2$ changes the solution
$v_0/\Lambda$ of Equation (\ref{eq:gap}) and hence the ratio
$m_t/\Lambda$. As we said, choosing the values of the bare parameters
completely determines everything in the theory. For example, to attain
a very small value of $m_t/\Lambda$ requires $m_0^2/\Lambda^2$ to be
tuned quite precisely. Using Equation (\ref{eq:gap}), the critical
surface, where $v_0/\Lambda=0$, is the transition line
\be
\frac{m_0^2}{\Lambda^2} - \frac{N_F y_0^2}{4 \pi^2} = 0.
\label{eq:critical}
\ee
\FIGURE[ht]{
\includegraphics[width=7.2cm,height=5.5cm]{figs/flow_lam_version2.eps}
\includegraphics[width=7.2cm,height=5.5cm]{figs/flow_y_version2.eps}
\caption{\label{fig:largeN_RG} The exact RG flow of the renormalized 
couplings $\lambda$ and $y$ with
the full cutoff dependence. The corresponding bare couplings are
$\lambda_0=0.1$ and $y_0=0.7$. For large cutoff, the exact flow agrees
with the continuum RG flow, where the cutoff dependence is
omitted. For small cutoff, the exact RG flows to the bare couplings
$\lambda_0$ and $y_0$, but the continuum RG misleadingly predicts
that $\lambda$ turns negative.}
}
Using Equations (\ref{eq:barerenorm}) and (\ref{eq:solve2}), all of the
counterterms and renormalized parameters can be expressed in terms of
$\lambda_0, y_0, m^2_0$ and $v_0$. Solving this
set of simultaneous equations is a simple numerical exercise. We make an
arbitrary choice $\lambda_0=0.1, y_0=0.7$ which would correspond to the physical
Higgs below its lower bound in phenomenological considerations. Varying the value of
$m_0^2/\Lambda^2$, we explore numerically the range $10^{-13} < m_t/\Lambda <
10^2$. The results in a limited range are plotted in Figure \ref{fig:largeN_RG}. 
When the cutoff is
high, the exact RG flow is exactly the same as if the cutoff had been
completely removed and follows precisely the continuum form of
Equation (\ref{eq:contflow}). However, as the cutoff is reduced, the
exact RG flow eventually breaks away from the continuum form and
reaches a plateau at the value of the bare coupling. 

The continuum RG in the above example predicts that $\lambda$ turns 
negative at some energy scale as the flow
continues. This was used in the past as an indication that the
ground state of the theory turns unstable at that scale which would
determine the energy scale of new physics necessary to sustain
a particular value of the physical Higgs mass (vacuum instability bound).
As shown above, the true RG flow with the full cutoff dependence saturates at
$\lambda_0$ and does not turn negative under the necessary $\lambda_0 > 0$
stability requirement of the model. This makes the phenomenological RG method 
and the apparent vacuum instability quite
suspect in the presence of the non-removable finite
cutoff which is required by triviality of the renormalized couplings.

The absence of vacuum instability will
be demonstrated directly in the next section using the Higgs effective potential.
In sections 4 and 5 we will propose a lattice strategy to determine the Higgs mass lower
bound in the presence of an intrinsic cutoff without relying on the continuum RG flow.
In this new strategy even the $\lambda_0 > 0$ condition might be relaxed by adding new
irrelevant operators, like the $\frac{\lambda_6}{\Lambda^2} \phi^6$ term, to keep the
stability of the cutoff theory intact.

\section{The effective potential and vacuum instability}

First, we will present here the RG improved one-loop calculation of the effective potential 
with unstable vacuum when the cutoff is ignored. Next we show the absence of vacuum instability 
when the cutoff is correctly enforced. 

\subsection{Continuum 1-loop effective potential}

For the Higgs-Yukawa model with $N_F$ fermions of
Section 2, the 1-loop renormalized effective potential is
\bea
U_{\rm eff} &=& \frac{1}{2} m^2 \phi^2 + \frac{1}{24} \lambda \phi^4 +
\frac{1}{2} \delta m^2 \phi^2 + \frac{1}{24} \delta \lambda \phi^4 - 2
N_F \int_k \ln[ 1 + y^2 \phi^2/k^2] \nn
&+& \frac{1}{2} \int_k \left( \ln[k^2 + V''(\phi)] - \ln[k^2 + V''(0)]
\right), \nn 
V &=& \frac{1}{2} m^2 \phi^2 + \frac{1}{24} \lambda \phi^4,
\label{eq:Ueff1loop}
\eea
where the Higgs-loop contributions are also included now. For consistency,
we impose exactly the same renormalization conditions
Equations (\ref{eq:renorm2}) and (\ref{eq:renorm3}) used in Section 2,
including all the Higgs-loop radiative corrections. Because
$\delta y$ and $\delta z_\psi$ are non-zero (we no longer impose the
large $N_F$ limit), we specify the two additional renormalization
conditions. The fermion inverse propagator is
\bea
G^{-1}_{\psi \psi}(p) = p_\mu \gamma_\mu + y v + \delta z_\psi p_\mu
\gamma_\mu + \delta y v - \Sigma_F(p), \nn
\Sigma_F(p) = y^2 \int_k \frac{- k_\mu \gamma_\mu +
  yv}{(k^2 + y^2 v^2)((k-p)^2 + m_H^2)},
\label{eq:Gferm2}
\eea
the radiative correction coming from a single Higgs-loop diagram, and we
require that 
\be
G^{-1}_{\psi \psi}(p \rightarrow 0) = p_\mu \gamma_\mu + y v.
\label{eq:renormGferm}
\ee
This gives two renormalization conditions,
\bea
&&\delta y v - \Sigma_F(p \rightarrow 0)=0 ~, \nn
&&\delta z_\psi - \left. \frac{d \Sigma_F}{d (p_\mu \gamma_\mu)}
\right|_{p \rightarrow 0}.
\label{eq:renorm4}
\eea
Again, the counterterms completely remove all the finite and infinite parts
of the radiative corrections. We regulate the momentum integrals using
e.g.~a hard-momentum cutoff. The counterterms and the renormalized
effective potential are calculated exactly using a finite cutoff. We
then take the naive limit $\phi/\Lambda \rightarrow 0$ to remove all
cutoff dependence. This ignores the fact that a finite and possibly
low cutoff is required to maintain $\lambda, y \ne 0$ (a
crucial point why the instability does not occur in the presence of finite cutoff).

The continuum form of the 1-loop renormalized effective potential
is given by
\bea
U_{\rm eff} &=& \frac{1}{2} m^2 \phi^2 + \frac{1}{24} \lambda \phi^4 -
\frac{N_F y^4}{16 \pi^2} \left[ - \frac{3}{2} \phi^4 + 2 v^2 \phi^2 + \phi^4
  \ln \frac{\phi^2}{v^2} \right] \nn
&+& \frac{1}{16 \pi^2} \left[ \frac{1}{16} 
  ( \lambda^2 \phi^4 - 2 \lambda \phi^2 m_H^2 ) \ln \frac{m^2 + \lambda
  \phi^2/2}{m_H^2} \right. \nn
&& + \left. \frac{1}{16} m_H^4 \ln \frac{m^2 + \lambda
  \phi^2/2}{m^2} - \frac{3}{32} \lambda^2 \phi^4 + \frac{7}{16} \lambda
  \phi^2 m_H^2 \right],
\label{eq:fullUeff}
\eea
where $m_H^2=\lambda v^2/3$. Due to our choice of renormalization
conditions, the tree-level vev $v=\sqrt{3
  m_H^2/\lambda}$ is not shifted: one can check explicitly that $U_{\rm
    eff}$ in Equation (\ref{eq:fullUeff}) has its minimum at
$\phi=v$. The large $N_F$ limit can be recovered by omitting the
Higgs-loop terms.

\subsection{RG improved effective potential and vacuum instability}

The stability of the ground state is determined by
the behavior of $U_{\rm eff}$ for large $\phi$. We see from
Equation (\ref{eq:fullUeff}) that the dominant terms in this regime are
of the form $\lambda^2 \phi^4 \ln(\phi^2/v^2)$ and $-N_F y^4 \phi^4
\ln(\phi^2/v^2)$. The negative fermion term brings up the possibility
that the vev $v$ is unstable. Hence stability is determined by the
relative values of $\lambda^2$ and $y^4$, which are related to
$m_H$ and $m_t$. If the fermionic term dominates at large $\phi$, the
minimum at $v$ is only a local one and will decay. If we believe that the
vacuum is absolutely stable, then new degrees of freedom must enter at
the scale  where $U_{\rm eff}(\phi)$ first becomes unstable. For given
values of $m_H$ and $m_t$, this predicts the emergence of new
physics. Turning this around, let us fix $m_t$ and ask that no new
stabilizing degrees of freedom are needed for $\phi \le E$. Then we
obtain a lower bound  $m_H(E)$: if the Higgs is lighter than this,
$U_{\rm eff}$ is already unstable for $\phi$ below $E$ because the
fermion term dominates even earlier.

Improved vacuum instability can be shown via the running renormalized
couplings in RG setting. We can define a set of renormalization
conditions in the continuum, for example in the $\overline{MS}$ scheme, 
where the couplings flow with the renormalization scale $\mu$. The 1-loop RG
equations for the Higgs-Yukawa model are 
\bea
\mu\frac{dy}{d\mu} &=& \frac{1}{8\pi^2}(3 + 2 N_F)y^4, \nn
\mu\frac{d\lambda}{d\mu} &=& \frac{1}{16\pi^2}( 3 \lambda^2 + 8 N_F
  \lambda y^2 - 48 N_F y^4).
\label{eq:contflow1loop}
\eea
We can set the initial conditions $\lambda(\mu=v)=3 m_H^2/v^2$ and
$y(\mu=v)=m_t/v$. If $m_t$ is sufficiently heavy relative to $m_H$,
the Yukawa coupling dominates the RG flow and $d \lambda/d\mu <
0$. The renormalized Higgs coupling eventually becomes negative at
some $\mu=E$. If the instability occurs at very large $\phi/v$, large
logarithmic terms $\ln(\phi/v)$ in $U_{\rm eff}$ might spoil the
perturbative expansion. This can be reduced using renormalization
group improvement to resum the leading large logarithms. The dominant
terms of $U_{\rm eff}$ at large $\phi$ then become $\lambda(\mu)
\phi^4(\mu)$. Hence $\lambda(E)=0$ indicates that the ground state is
just about to become unstable.

\subsection{The constraint effective potential on the lattice}

We can calculate the exact effective potential non-perturbatively,
using lattice simulations. This was first shown in the pure Higgs
theory by Kuti and Shen \cite{Kuti:1987bs}.
There is some finite lattice spacing $a$ on the lattice which
restricts the momenta $|p_\mu| \le \pi/a$ replacing the sharp momentum
cutoff used in section 2. For a Higgs-Yukawa theory
with $N_F$ fermions, the Euclidean lattice partition function is
\bea
&& Z = \prod_x \int d\phi_0(x) [{\rm Det}(D[\phi_0])]^{N_F} \exp(-S[\phi_0]),
 =  \prod_x \int d\phi_0(x) \exp(-S_{\rm eff}[\phi_0]) \nn
&& S = \sum_x \frac{1}{2} m_0^2 \phi_0^2(x) + \frac{1}{24} \lambda_0
\phi_0^4(x) + \frac{1}{2}(\partial_\mu \phi_0(x))^2, \nn
&& (D[\phi_0])_{xy} = \gamma_\mu \partial_{\mu,xy} + y_0
\phi_0(x) \delta_{xy},
\label{eq:Z}
\eea
where the partial derivatives are replaced by finite
lattice differences. If the integrand is positive-definite, it can be
interpreted as a probability density and importance sampling
(i.e.~Monte Carlo integration) can be used to calculate expectation
values, e.g.~$\langle \phi_0 \rangle$, non-perturbatively with the exact
distribution $[{\rm Det}(D)]^{N_F}\exp(-S)$. 
All dimensionful quantities are calculated
in units of the lattice spacing $a$. There is a phase diagram in the
bare-coupling space $m^2_0, \lambda_0, y_0$. The Higgs phase and the symmetric
phase are separated by a second order transition, where the vev, $va$, and
the masses $m_Ha$ and $m_ta$, vanish. Since the vev and masses are
non-zero in physical units, the transition corresponds to the
continuum limit $a\rightarrow0$. To make the cutoff $\Lambda=\pi/a$
large, the bare couplings must be tuned to be close to the transition
line. If we calculate via simulations that e.g.~$av=\langle a \phi
\rangle \approx 0.05$ for some choice of bare couplings, we can use
$v=246$~GeV to convert this into a cutoff $\Lambda \approx 15$~TeV, as
well as determine $m_H$ and $m_t$ in physical units.

In a finite space-time volume $\Omega$, we will use the
constraint effective potential \cite{Kuti:1987bs,Fukuda:1974ey}. For a pure scalar
field theory, this is
\be
\exp(-\Omega U_{\Omega}(\Phi)) = \prod_x \int d\phi(x) \delta\left(\Phi -
\frac{1}{\Omega} \sum_x \phi(x)\right) \exp(-S[\phi]).
\label{eq:constraintdef}
\ee
The delta function enforces the constraint that the scalar field $\phi$
fluctuates around a fixed average $\Phi$. The constraint effective
potential $U_{\Omega}(\Phi)$ has a very physical interpretation. If
the constraint is not imposed, the probability that the system
generates a configuration where the average field takes the value
$\Phi$ is
\be
P(\Phi) = \frac{1}{Z}\exp(-\Omega U_{\Omega}(\Phi)), \hspace{5mm}
Z = \int d\Phi' \exp(-\Omega U_{\Omega}(\Phi')).
\label{eq:constraintprob}
\ee
This is in very close analogy to the probability distribution for the
magnetization in a spin system. The scalar expectation value $v=\langle
\phi \rangle$ is the value of $\Phi$ for which $U_\Omega$ has an
absolute minimum. In a finite volume, the constraint
effective potential is non-convex and can have multiple
local minima \cite{O'Raifeartaigh:1986hi}. The standard effective
potential $U_{\rm eff}(\Phi)$ is always convex, even in a finite
volume, as the Maxwell construction connects the various minima. The
two effective potentials are identical in the infinite-volume limit,
$\lim_{\Omega \rightarrow \infty}U_{\Omega}(\Phi)=U_{\rm eff}(\Phi)$,
and the constraint effective potential recovers the convexity
property. In a finite volume, it is more useful to work with the
constraint effective potential, where multiple minima can be observed
and the transition between the Higgs and symmetric phases is clear. It
is also more natural, as the probability distribution $P(\Phi)$ can be
directly observed in lattice simulations. For the rest of this paper,
we drop the subscript $\Omega$.

\subsection{Hybrid Monte Carlo algorithm and the effective potential}
One way to extract the effective potential from lattice simulations
is to generate the ensemble of configurations, calculate the average
scalar field $\Phi$ for each configuration and hence the
probability distribution $P(\Phi)$. The effective potential is
extracted by numerically fitting $U_{\rm eff}(\Phi)$ to $P(\Phi)$
using Equation (\ref{eq:constraintprob}). This gives
the effective potential for all $\Phi$ from one simulation, but with
limited accuracy. An alternative method is calculate the derivative of
the effective potential. For the Higgs-Yukawa model with $N_F$
degenerate fermions, the derivative is
\be
\frac{d U_{\rm eff}}{d \Phi} = m^2 \Phi + \frac{1}{6} \lambda
\langle \phi^3 \rangle_{\Phi} - N_F y \langle \bar{\psi} \psi
\rangle_{\Phi}, \hspace{5mm}
\langle \bar{\psi} \psi \rangle_{\Phi} = \langle {\rm Tr}
(D[\phi]^{-1}) \rangle_\Phi.
\label{eq:potderiv}
\ee
The expectation values $\langle ... \rangle_{\Phi}$ mean that, in the
lattice simulations, the scalar field fluctuates around some fixed
average value $\Phi$. This method determines the effective potential
with greater accuracy than fitting the distribution $P(\Phi)$, but the
drawback is that a separate lattice simulation has to be run for every
value of $\Phi$. This is the method we use in our investigation
of the vacuum instability.

In this section we use staggered fermions \cite{Susskind:1976jm,
 Kogut:1974ag}, one flavor of which corresponds to four
fermion flavors in the continuum. With one staggered fermion, the
determinant ${\rm Det}(D)$ is real but can be negative due to $\phi$
fluctuations. Then the partition function integrand is not
positive-definite and Monte Carlo integration cannot be applied. To
overcome this problem, we simulate two staggered fermions,
corresponding to eight continuum flavors, as $[{\rm Det}(D)]^2$
guarantees a positive-definite density.We used staggered fermions only in the
very early phase of our simulations. The complicated taste structure of staggered
fermions with the related rooting issues and the lack of full chiral symmetry motivated
the switch to chiral overlap fermions which are used now exclusively in our Higgs project. 
Staggered results for the effective potential, which are used here mainly for simplicity and
pedagogy, have been replaced by simulations with chiral overlap fermions.

Configurations are generated
using the Hybrid Monte Carlo algorithm \cite{Duane:1987de}, where a
fictitious time $t$ and momenta $\pi(x,t)$ are introduced. New
configurations are generated from the equations of motion
\bea
&& \dot{\phi}(x,t) = \pi(x,t)~, \nn
&& \dot{\pi}(x,t) = - \left[ \frac{\partial S_{\rm eff}}{\partial
\phi(x,t)} - \frac{1}{\Omega} \sum_y \frac{\partial S_{\rm eff}}
{\partial \phi(y,t)} \right],
\label{eq:HMC}
\eea
where the effective action $S_{\rm eff}$ is given in
Equation (\ref{eq:Z}). The second term in $\dot{\pi}(x,t)$ is included
to enforce the constraints
\be
\frac{1}{\Omega} \sum_y \phi(y,t) = \Phi, \hspace{5mm} \sum_y \pi(y,t)
= 0.
\ee
We work with fixed lattice volumes of size $8^3 \times 16$. The scalar
field has periodic boundary conditions, the fermionic field is periodic
in the short directions and antiperiodic in the long direction. We use the
standard leapfrog method to solve the equations of motion, where the
step-size $\Delta t$ is adjusted to achieve acceptance rates well
above 90\%, and each trajectory length satisfies $N_t \Delta t \ge
1$. For each simulation, we generate at least $10^4$ configurations
and check that correlations between the configurations are small.

The basic quantities of the theory are the bare fields and
couplings. A particular choice of bare couplings puts us somewhere in
the phase diagram and all physical quantities are now fixed. A
separate constrained simulation is run for each value of $\Phi_0$ to
calculate the effective potential derivative. The expectation values
we measure on the lattice are bare ones, so the simulations give the
bare equivalent of Equation (\ref{eq:potderiv}), namely
\be
\frac{d U_{\rm eff}}{d \Phi_0} = m_0^2 \Phi_0 + \frac{1}{6} \lambda_0
\langle \phi_0^3 \rangle_{\Phi_0} - N_F y_0 \langle \bar{\psi_0} \psi_0
\rangle_{\Phi_0}, \hspace{5mm}
\langle \bar{\psi_0} \psi_0 \rangle_{\Phi_0} =
\langle {\rm Tr} (D[\phi_0]^{-1}) \rangle_{\Phi_0},
\label{eq:potderivbare}
\ee
which is converted using the relationship between the bare and
renormalized fields,
\be
\Phi = \frac{\Phi_0}{\sqrt Z_\phi}, \hspace{5mm}
\frac{d U_{\rm eff}}{d \Phi} = \sqrt Z_\phi \frac{d U_{\rm eff}}
{d\Phi_0}.
\label{eq:barephi}
\ee
We measure the wave function renormalization factor $Z_\phi$ in
separate unconstrained simulations.

We want to follow the behavior of $U_{\rm eff}$ as we approach the
continuum limit, the critical surface in the bare-coupling space.
We make an arbitrary choice $y_0=0.5$ and $\lambda_0=0.1$. The
distance from the continuum limit is determined by the remaining bare
coupling $m_0^2$. We obtained results for three choices
$m_0^2=0.1,0.25$ and $0.29$. Typical non-perturbative measurements of the
derivative $dU_{\rm eff}/d\Phi$ are shown in Figure \ref{fig:Ueff1}. 
All dimensionful quantities
are in lattice units, e.g.~$a\cdot\Phi$. What do we expect to see? In
the Higgs phase of the theory, $U_{\rm eff}$ should have a
local maximum at the origin and a local minimum for some non-zero
$a\cdot\Phi$. If the vacuum is stable, the local minimum is
in fact an absolute one. Let us first look at the results for
$m_0^2=0.1$, shown in Figure \ref{fig:Ueff1}. 
The simulations show that
$dU_{\rm eff}/d\Phi$ vanishes at the origin and at $a\cdot\Phi \approx
2.0$; these are the extrema. The derivative is negative between these
points, so the origin is indeed a local maximum.
For $a\cdot\Phi > 2$, the
derivative is always positive and the local minimum appears to be an
absolute one. If the vacuum is unstable, $dU_{\rm eff}/d\Phi$ should turn
negative at large $a\cdot\Phi$, for which the simulations show no
evidence. In these units, the lattice cutoff is $\Lambda = \pi/a$ and
the ratio of cutoff to scalar expectation value is $\Lambda/v \approx
1.5$. This is far from the continuum limit. 
\vskip 0.4cm
\FIGURE[h]{
\includegraphics[width=6.4cm]{figs/Ueff_cont_latt_simul_1_latest.eps}
\phantom{.}\vspace*{0.1cm}
\includegraphics[width=6.1cm]{figs/Ueff_cont_latt_simul_2_latest.eps}
\caption{\label{fig:Ueff1}
The derivative of the effective potential $dU_{\rm eff}/d\Phi$ for
the bare couplings $y_0=0.5, \lambda_0=0.1, m_0^2=0.1$, for which the
vev is 
$av=2.035(1)$.
The left side plot is a close-up of the behavior
near the origin. The circles are the results of the simulations and
the curves are given by continuum and lattice renormalized
perturbation theory.}
}

We vary the bare mass to get closer to the critical surface and
the continuum limit for
bare masses $m_0^2=0.25$ and 0.29 respectively. The simulations
show the same qualitative behavior for $U_{\rm eff}$: the origin is a
local maximum, there is an absolute minimum for some non-zero $a\Phi$
and no sign of an instability in the potential. The minimum occurs at
$a\Phi \approx 0.81$ and $0.47$ respectively, for which $\Lambda/v
\approx 3.9$ and 6.7, pushing towards the continuum limit.

The first check of these calculations is to run separate unconstrained
simulations with the same bare couplings, where $\sum_x \phi(x)$ is
allowed to fluctuate freely, and to measure independently $v=\langle \phi
\rangle$. This expectation value should be identical to the value of
$\Phi$ where $U_{\rm eff}$ has an absolute minimum, as determined by
the constrained simulations. In the unconstrained simulations, the
second term for $\dot{\pi}(x,t)$ in Equation (\ref{eq:HMC}) is omitted. The
results of the unconstrained simulations are given in Table
\ref{table:Zvevmhmt}. There is indeed perfect agreement
between the measurements of $\langle a\phi \rangle$ and the location
of the $U_{\rm eff}$ minimum obtained from the constrained
simulations.
\TABLE[t]{
\begin{tabular}{ccccccc}
\hline \hline
$y_0$ & $\lambda_0$ & $m_0^2$ & $Z_\phi$ & $av=\langle a \phi \rangle$
& $a m_H$ & $a m_t$ \\
\hline \hline
0.5 & 0.1 & 0.1 & 0.987(1) & 2.035(1) & 0.521(5) & 0.9977(5) \\
 & & 0.25 & 0.9705(8) & 0.811(1) & 0.297(4) & 0.3906(7) \\
 & & 0.29 & 0.9676(7) & 0.4685(6) & 0.248(3) & 0.2230(3) \\
\hline
\end{tabular}
\caption{{} The wave function renormalization factor, the renormalized
  scalar expectation value and the  Higgs and Top masses, obtained from
  unconstrained lattice simulations. The bare couplings are those used
  for the lattice measurements of the effective potential $U_{\rm
    eff}$. The estimated errors are in parentheses.}
\label{table:Zvevmhmt}
}
The continuum perturbation theory calculation of $U_{\rm eff}$ is also
shown in Figure~\ref{fig:Ueff1}. We only display the large $N_F$ result: not
surprisingly, for $N_F=8$, the Higgs-loop contributions are negligible
and can be omitted. We see excellent agreement with the
non-perturbative simulations for $\Phi \lesssim v$, as shown in the
left side plot. However, the behavior as $\Phi$ increases is completely
different, as shown in the right side plot. Continuum perturbation theory
breaks away from the simulation results and predicts that the vacuum
becomes unstable, with $dU_{\rm eff}/d\Phi$ turning negative. The exact
non-perturbative calculation shows no indication of this. 

What can we conclude from the comparison? Continuum perturbation
theory works well for $\Phi$ less than and even close to the lattice
cutoff $\Lambda=\pi/a$, as shown by the very good agreement with the
exact lattice calculations. This is the most that one could have
expected. The instability is predicted at $\Phi$ well above the
cutoff, which is completely unphysical and where one cannot expect the
continuum calculation to apply. The exact effective potential, with
the full cutoff dependence, is absolutely stable. The standard
interpretation of the instability in the continuum $U_{\rm eff}$ would
be to say new physics appears at this energy scale to stabilize the
ground state. But the actual cutoff of the field theory is far below
this scale, especially as we get closer to the continuum limit. The
instability only appears when the finite cutoff effects are ignored
--- there is no need for new physics. One can ask, is it possible to
arrange both the standard ground state and the instability to occur
well below the regulator cutoff? If so, the instability would be a
genuine low-energy prediction. The answer is no
in the Top-Higgs Yukawa model, if only the standard terms are included
in the lattice Lagrangian. In this case the only freedom one
has is the choice of the bare couplings, and nowhere in the
coupling-space is a genuine instability seen. If higher dimensional operators
are included, the $\lambda_0 > 0$ condition perhaps could be relaxed by adding new
irrelevant operators, like the $\frac{\lambda_6}{\Lambda^2} \phi^6$ term, to keep the
stability of the cutoff theory intact. This scenario requires further investigation.

It can be shown in renormalized lattice perturbation theory 
that the breakdown of continuum perturbation theory is due solely to the finite
cutoff. A finite cutoff is used in the lattice momentum
integrals for the radiative corrections and the counterterms of $U_{\rm eff}$, but
otherwise the procedure is the same as in the continuum.
In Figure \ref{fig:Ueff1} we see excellent
agreement between simulations, and lattice and continuum renormalized
perturbation theory for $\Phi/v \lesssim 1$. As
$\Phi$ increases, lattice perturbation theory {\em exactly} tracks the
non-perturbative result, showing a perfectly stable ground state. The
continuum calculation breaks down, not because of large couplings, but
because of the neglected finite cutoff.

\section{Wilsonian renormalization group and vacuum instability}

Most of the original work on
the consistency of quantum field theory considered only idealized
theories, supposedly fundamental 
to describe physics at arbitrarily high energies. Although in the previous section
on vacuum instability and the related Higgs lower bound problem we found a non-removable
intrinsic cutoff, the analysis was based on the traditional renormalization procedure.
The Wilsonian viewpoint of the renormalization group provides a broader and more
complete perspective on the discussion.

\subsection{Wilson's running Lagrangian}

In the 1970s Wilson developed a new, intuitive way of looking at the 
renormalization of quantum field theories based on the flow of effective Lagrangians as
generated by renormalization group transformations~\cite{Wilson:1973jj}. 
This is based on the
realization that physics as we know it seems to be described by
effective quantum field theories, which are useful only up to the energy
scale $\Lambda_0$ where new and yet unknown physics is reached.
Some smooth intrinsic regularization is introduced (inherited from new UV physics) 
at $\Lambda_0$
which in Euclidean space restricts the length $p^2$ of all four--momenta.
Physics below the cutoff scale $\Lambda_0$ is described by a very general
`bare' Lagrangian ${\cal{L}}(\Lambda_0)$ with an infinite series of
local terms, constrained only by symmetries. For any choice of the
coupling constants in the local terms of the bare Lagrangian, the
Euclidean path integral of the partition function has to be finite and
well defined.
The most fundamental constraint on
the bare Lagrangian is the existence and stability of the functional integral which defines the
Euclidean partition function.
If the viewpoint of `naturalness' is adopted, all the coupling constants of the
higher dimensional operators are chosen to be of order
one in units of $\Lambda_0$. 
Using Wilson's exact
renormalization group we can consider smoothly lowering the
regularization scale
to some value $\Lambda_R$ say, of order the energy scale $E$ 
far below $\Lambda_0$. To
keep physics unchanged, the coupling constants must change with
the regularization scale. Hence we have a running, or effective
Lagrangian ${\cal{L}}(\Lambda)$, which flows with $\Lambda$ and remains
stable at every stage of the procedure in the sense of a convergent
Euclidean path integral.
Since we can use the Lagrangian ${\cal{L}}(\Lambda_R)$
to calculate low energy physics at the scale $E$, it is 
not the coupling constants at $\Lambda_0$ that are
important, but those at the scale $\Lambda_R$. The bare couplings have to
be close to a critical surface if $m_{ph}/\Lambda_0 \ll 1$
for the low energy physical masses $m_{ph}$ of the theory.

An effective field theory is renormalizable if we can calculate
all the S--matrix elements for processes with energy scale
$E$, up to small errors which vanish as powers of $E/\Lambda_0$, once
we have determined a finite number of coupling constants at some
renormalization scale $\Lambda_R \sim E$. These coupling constants
are called relevant; all others are irrelevant.
Whatever values we choose for $\Lambda_0$ (as long as it is large enough)
and the irrelevant bare couplings $\eta(\Lambda_0)$
(as long as they are natural enough),
for a particular choice of
the relevant operator set $\lambda(\Lambda_R)$, 
the irrelevant operator set $\eta(\Lambda_R)$ will be of the order of some
power of $(\Lambda_R/\Lambda_0)$.
In other words, for any point on the submanifold of relevant couplings 
at $\Lambda_R$ there
is a flow towards it from a wide variety of initial Lagrangians at
$\Lambda_0$, all of these being equivalent as far as
the values of S--matrix elements for processes with energies of order
$E\sim\Lambda_R$ are concerned. This more general aspect of
renormalizability is the concept of universality.
An effective quantum theory thus gives us a much more general notion
of renormalizability than we had in conventional quantum field theory:
the regularization need no longer be removed, and the irrelevant bare
couplings need not be zero.

It is useful now to adopt the Wilsonian view on the running effective
Lagrangian to the Top-Higgs Yukawa model we investigated in the previous section.

\subsection{Top-Higgs Yukawa model, vacuum instability, and running Lagrangian}

Adapting the notion of the the running Wilson Lagrangian for the Top-Higgs Yukawa
model, there are only two marginally irrelevant couplings, $\lambda(t)$ and $y(t)$,
in addition to the relevant Higgs mass operator. It is important to note that the 
couplings for increasing $t={\rm log}(\Lambda_0/\Lambda)$ flow from bare 
$\lambda_0$ and $y_0$ toward
their low energy renormalized values as a function of the energy scale. For example,
in the large $N_F$ limit and for large
$t$ values, neglecting the irrelevant couplings, the flows are expected to look
approximately the same as described by Equation~(\ref{eq:contflow}). The Yukawa coupling
$y(t)$ will monotonically decrease from its bare value $y_0$ towards zero,
at the logarithmic rate
of Equation~(\ref{eq:yvanish}) for large $t$. 
The Higgs coupling will start from its bare value
$\lambda(0)=\lambda_0$ and either it will monotonically decrease, or
after some initial rising it will turn around and continue to decrease
monotonically towards zero,
at the logarithmic rate of Equation~(\ref{eq:lamvanish}) for large $t$. In the
Wilsonian picture, all RG trajectories flow from the general coupling
constant space of cutoff Lagrangians
${\cal{L}}(\Lambda_0)_{Top-Higgs}$ towards the trajectory specified by 
(\ref{eq:yvanish}) and (\ref{eq:lamvanish}) with small but calculable corrections
from irrelevant operators in the large $t$ limit. 

In the Wilsonian view of the running Lagrangian, the cutoff dependent
Higgs mass lower bound can be determined in the space of the bare 
cutoff Lagrangians ${\cal{L}}(\Lambda_0)_{Top-Higgs}$ from the smallest allowed value of $\lambda(\Lambda_R)$ for a fixed 
$\Lambda_0/\Lambda_R \ll 1$ ratio where a natural choice for $\Lambda_R$ is the weak
boson mass $m_Z$, or the vacuum expectation value $v$.
This calculation is, of course,
very hard to implement operationally with a large number of bare couplings. The important
stability condition is the only constraint (with, or without naturalness) on the
space of cutoff Lagrangians.
For example, the choice of $\lambda_0 < 0$ a priori should not be excluded 
at the cutoff scale $\Lambda_0$, 
but it requires the
presence of some positive higher dimensional operator, like 
$\lambda_0^{(6)}/\Lambda^2\cdot \phi^6$, with $ \lambda_0^{(6)} > 0$, to provide stability.
Whether the Higgs mass lower bound will be necessarily associated with the limit 
$\lambda_0\rightarrow 0$, or the $\lambda_0 < 0$ region also needs to be explored remains an unresolved and interesting question.

In phenomenological applications an attempt is always 
made to simplify Wilson's framework
of dealing with the full space of running Lagrangians. Invoking the 
$\Lambda_0/\Lambda_R \rightarrow 0$ limit, only the running of the relevant and marginally irrelevant couplings
is calculated and the effects of irrelevant operators are ignored. In addition, in the application
of RG equations to the vacuum instability problem, the simplified equations on $\lambda(t)$ and $y(t)$ are running backward from the $m_Z$ scale towards the cutoff $\Lambda_0$. 
This interchange of
the natural Wilsonian UV $\rightarrow$ IR flow with the IR $\rightarrow$ UV integration of
relevant couplings only
is a nontrivial proposition because the Wilsonian RG flow is not known to be reversible, 
and to set all the irrelevant couplings to zero at the scale $\Lambda_R=m_Z$
would require an unknown extension of the space of cutoff Lagrangians
${\cal{L}}(\Lambda_0)_{Top-Higgs}$, if it exists at all. 

In most of the 
phenomenological RG applications this is not a problem. We believe, however, that the RG 
treatment of the vacuum instability problem requires special care. What corresponds to the 
unstable $U_\text{eff}$ in Figure~\ref{fig:Ueff1} is the running $\lambda(t)$ which at some scale $t_0$,
far below the cutoff scale, turns negative as the RG is running backward, from
$t={\rm log}\Lambda_0/m_Z$ towards the cutoff scale $t=1$. It is a signal
that higher dimensional operators must play a role to provide a continued stability to 
the theory on all scales. It is unlikely that a positive $\lambda_0$ on the cutoff scale 
can support this picture, forcing the running $\lambda (t)$ to turn positive again and produce
an effective potential which will turn back positive again after a second minimum which might
be lower than the original one where the spontaneously broken theory was built (decay of the
false vacuum). It is more likely that this scenario, if it exists at all, will require 
the $\lambda_0 < 0$ extension of the space of bare Lagrangians. This is an extension which remains
largely unexplored and we are just beginning to investigate it.

\subsection{Phenomenology from 2-loop continuum RG}

Vacuum instability was first raised in \cite{Krive:1976sg}
and it has since been increasingly refined in application to the
Standard Model \cite{Linde:1977mm, Politzer:1978ic, Cabibbo:1979ay, 
Hung:1979dn, Flores:1982rv, Lindner:1985uk, Sher:1988mj, Lindner:1988ww,Ford:1992mv, 
Sher:1993mf, Casas:1994qy, Altarelli:1994rb,
Casas:1996aq, Boyanovsky:1997dj, Einhorn:2007rv}.
The state-of-the-art calculation determines the
effective potential to one-loop order, with RG improvement applied up
to two-loop order to the running couplings.
\FIGURE[h]{
\includegraphics[width=8cm,angle=0]{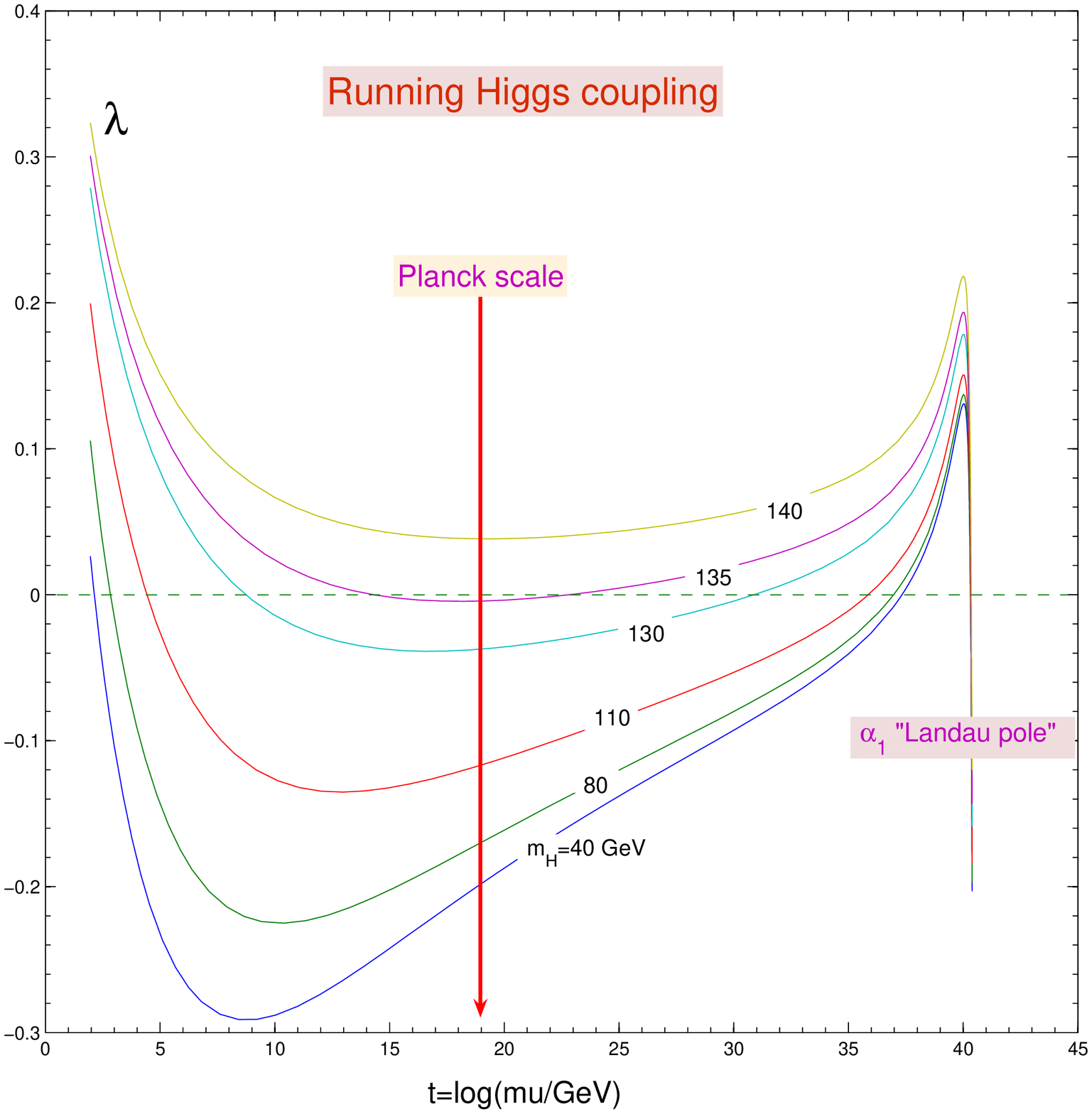}
\caption{\label{fig:RG_2loop} The running Higgs coupling
is plotted for different choices of the Higgs mass
from our numerical solution of the five 
coupled 2-loop RG equations for the $\lambda,y,g_1,g_2,g_3$ couplings.
For input, $m_t=175$ GeV was used with the experimental values of the 
$g_1,g_2,g_3$ gauge couplings. The 1-loop matching of the couplings and
the starting scale of the RG was chosen at $m_Z$.
} 
}
\noindent Results from
\cite{Casas:1996aq} exhibit the unstable Standard Model effective potential for
$m_H=52$~GeV and $m_t=175$~GeV, where the instability appears at
$\phi=1$~TeV. The lower bound shown in Figure \ref{fig:PDG} is also
taken from \cite{Casas:1996aq}. The finite width of the lower bound is an
estimate of the uncertainty of the theoretical calculation, including
the effect of unconsidered higher-order contributions. 
The strict lower bound for the Higgs mass can be further refined if one allows the ground
state to be unstable, but demands that the time required to tunnel
away from the local minimum at $v = 246$~GeV is longer than the lifetime
of the universe \cite{Arnold:1989cb, Anderson:1990aa, Arnold:1991cv,
Espinosa:1995se, Isidori:2001bm}. 

It is clear that the current
experimental limits on $m_H$ bring the lower bound into play. For
example, a Higgs boson with a mass of 100~GeV should indicate a
breakdown of the Standard Model around 50~TeV. However, a Higgs mass in
the range 160 -- 180~GeV apparently allows the Standard Model to be
valid all the way up to the Planck scale. 
The occurrence of the vacuum instability mostly relies on the relative
magnitudes of $\lambda^2$ and $y^4$ while both renormalized
couplings can remain small and all three gauge couplings of the SM are included.
The perturbative
RG approach, if cutoff effects can be safely ignored, seems to be on solid footing.
However, cutoff effects played an important role in Top-Higgs Yukawa models
where only the Higgs coupling $\lambda$ and Yukawa
coupling $y$ drive the dynamics. In this approximation we have shown
that vacuum instability cannot be induced with the SM Higgs potential
in the cutoff Lagrangian (the possible role of higher dimensional operators to induce
vacuum instability remains unclear, as we noted earlier).
However, in the phenomenological
application, all five couplings are running and it is important to ask: for the cutoff
$\Lambda$ at or below the Planck scale $M_P$, should we expect Top quark induced vacuum
instability with the SM cutoff Lagrangian without adding new operators? Do we expect
a qualitatively different picture when compared to the Top-Higgs Yukawa model?
From Figure~\ref{fig:RG_2loop} we find that the running $\lambda$ turns negative
below the Planck scale for Higgs mass values lower than 135 GeV and remains negative when
$M_P$ is reached. Further lowering the Higgs mass lowers 
the scale where $\lambda$ turns negative.
It remains unclear how these RG flows would be effected by holding $\lambda_0 > 0$ 
in the SM Higgs Lagrangian at some cutoff scale $\Lambda$. 
How some higher dimensional operators might provide a well-defined cutoff
theory for the choice $\lambda_0 < 0$ will require further investigation.

\section{Higgs mass lower bound from the lattice}

We would like to outline and implement the first step of a robust strategy 
to calculate the lower Higgs
mass bound as a function of the lattice momentum cutoff. The question about 
breaking Euclidean invariance with the lattice cutoff will eventually have to be addressed also.

\subsection{Yukawa couplings of the Top and Bottom quarks}

The third, heaviest generation of quarks consists of the
left-handed $SU(2)$ top-bottom doublet
$
Q_{{\rm L}}=\binom{t_L}{b_L}
$
and the corresponding right-handed $SU(2)$ singlets $t_{{\rm R}},\,b_{{\rm
    R}}$. The complex $SU(2)$ doublet Higgs field
$\Phi(x)$ with $U(1)$ hypercharge $Y=1$ is $\Phi=\binom{\phi^{+}}{\phi^0}$
where the suffixes +,0 characterize the electric charge +1, 0 of the
components. Since $\phi^{+}$ and $\phi^{0}$ are complex, we can introduce four
real components, $\Phi=\binom{\phi_{1}+i\phi_{2}}{i\phi_{3}+\phi_{4}} $
and the Higgs potential will have O(4) symmetry, with broken custodial O(3) symmetry, if
the Yukawa couplings $y_t$ and $y_b$, defined below, are different.
The Higgs potential in the complex doublet notation 
has the form,
\begin{equation}
  V(\Phi)=\frac{1}{2}m^2\Phi^{\dag}\Phi+\frac{\lambda}{24}(\Phi^{\dag}\Phi)^2.
\label{eq:higgs_pot}
\end{equation}
The Higgs field acquires a vacuum expectation value responsible for the
spontaneous electroweak symmetry breaking with $\langle\phi_4\rangle=v$ and the first three
components vanishing. The vacuum expectation value $v$ can be related to the Higgs
coupling constant by $v=\sqrt{3/\lambda}m_H$
with the relation between the Higgs mass $m_H$ and $m$ 
given by Equation~(\ref{eq:mhiggs}).

Of the four Higgs components three represent Goldstone degrees of
freedom, which at finite weak gauge coupling become the longitudinal 
degrees of freedom of the massive
weak gauge bosons with mass $m_W=vg_2/2$. The fourth component 
corresponds to the physical Higgs boson field.
We do not use the Higgs mechanism in the limit of zero weak gauge couplings
and keep all four Higgs field components where the $\phi_1,\phi_2,\phi_3$
fluctuations represent Goldstone particles with the symmetry breaking in
the $\phi_4$ direction.
In the SM Lagrangian all four Higgs components are treated on
equal footing where 
${\cal L}_{{\rm Yukawa}}$ describes the interactions of the $SU(2)_L$
doublet Higgs field with the quark fields
\begin{equation}
  {\cal L}_{{\rm Yukawa}}=y_t\cdot \overline{Q}_{{\rm L}}\Phi^{c}
  t_{R} + y_b\cdot \overline{Q}_{{\rm L}}\Phi b_{R} +{\rm h.c.}
\label{eq:top1}
\end{equation}
$\Phi^{c}=i\tau_2\Phi^{*}$ is the charge conjugate of $\Phi$, $\tau_2$ the
second Pauli matrix, $y_t,\,y_b$ are the top and bottom Yukawa couplings,
respectively. When they are equal, the O(3) custodial symmetry of the 
Higgs potential is preserved after symmetry breaking. For unequal couplings, only the $SU(2)_L$ symmetry of the Lagrangian is maintained. 
It is easy to write out the Yukawa couplings in components:
\begin{eqnarray}
{\cal L}_{{\rm Yukawa}}&=&y_t\{ \overline{t}_L (\phi_4 -i\phi_3)t_R +
 \overline{b}_L (i\phi_2-\phi_1)t_R\} + \\ \nonumber
&&y_b\{ \overline{t}_L (\phi_1+i\phi_2)b_R +
 \overline{b}_L (i\phi_3+\phi_4)b_R\} ~ + ~h.c.
\label{eq:top2}
\end{eqnarray}
All masses are proportional to $v$ as they are induced by spontaneous symmetry breaking.

\subsection{One-component Top-Higgs Yukawa model}

We have used lattice simulations to study the Higgs-Yukawa model
with a single real scalar field coupled to
the Top quark with three colors using  chiral overlap fermions.
This theory has only a Higgs particle and no Goldstone
bosons, and the Top quark color
indices correspond to three degenerate fermions. We will not be able to
calculate a lower bound directly relevant to phenomenology. Our
purpose here is to explain in a simpler model how this
non-perturbative calculation can be applied to a more realistic
approximation of the Standard Model.

The Yukawa interaction Lagrangian in Equation~(\ref{eq:top2}) has
a straightforward chiral lattice implementation in
the overlap formulation where the chiral left-handed and right-handed 
fermion components are precisely defined. The simulation of the full
doublet with the heavy Top and much lighter b quark would be very difficult 
on the lattice with two very different mass scales for $m_t$ and $m_b$ after
spontaneous symmetry breaking. 

One could choose for a pilot study the degenerate
case $y_t=y_b$ which has a recent 
lattice implementation~\cite{Gerhold:2007gx,Gerhold:2007yb}. In this limit, there are
three massless Goldstone particles contributing to Top-Higgs dynamics. When the weak
gauge couplings are turned on, the massless Goldstone modes 
become the longitudinal components of the massive weak gauge bosons 
via the Higgs-Kibble mechanism. The limitation of the four-component model 
with degenerate quark doublet is the artificially
enhanced fermion feedback into Higgs dynamics. 

Although the degenerate model of the Top and Bottom quarks is easy
to accommodate in our Higgs lattice toolbox, we chose the single component Higgs
Yukawa model for our pilot study with only the Top quark included. 
When the weak gauge couplings are turned on, one can choose unitary gauge
to eliminate the three Goldstone components. In this gauge, ignoring the weak gauge coupling
effects to leading order, one is left with diagonal Top and Bottom quark Yukawa couplings where the
b quark is decoupled in the $y_b=0$ limit. This is not a full justification for keeping the single
Higgs field only,
and the price to pay is the absence of feedback from the Goldstone modes into Higgs
dynamics. Since the primary purpose of the initial phase of our Higgs project is to develop
a comprehensive Higgs lattice toolbox and test its various uses, the limited 
one-component Higgs field dynamics will provide very useful information. The next logical step
will be to restore the four components of the Higgs field which requires the
b quark, and break the mass degeneracy moving
toward the $y_b \ll y_t$ limit.

\subsection{Phase diagram with chiral overlap fermions}

Lattice Yukawa models with staggered and Wilson fermions were studied 
before~\cite{Lin:1990ue,Bock:1990tv,Lee:1989mi}.
In this work, we adopted the overlap fermion operator to
represent the chiral Yukawa coupling between the Top quark fermion field
and the Higgs field. Although this is the most demanding choice for dynamical fermion
simulations, staggered and Wilson fermions are not suitable for
our goals. We discussed some difficulties with staggered fermions in section 3. The difficulties
with Wilson fermions are worse. It turns out to be impossible to tune to the critical surface
of the Top-Higgs lattice Yukawa model with Wilson fermions while keeping the Wilson doublers
on the cutoff scale. This is different from QCD applications of Wilson fermions. 
 
Our massless overlap Dirac operator is defined as 
$ a\cdot D =  1 + \gamma_5 \rm{sign}(H_w )$
with $H_w=\gamma_5 D_w$ where $D_w$ is the usual Wilson-Dirac
matrix with a negative mass which for $a=1$ has the form
\be
(D_w)_{yx} = 3\delta_{xy} - 
\frac{1}{2}\sum_\mu\biggl( (1+\gamma_\mu)U_\mu(x-y)\delta_{x,y+\mu}  +
(1-\gamma_\mu)U^{\dagger}_\mu(x)\delta_{x,y-\mu}\biggr) ~.
\label{eq:D_w}
\ee
Using the modified $\hat{\gamma}_{5}=\gamma_{5}(1-aD)$ gamma matrix,
we define two projection operators,
$
P_{\pm} = \frac{1}{2}( 1 \pm \gamma_{5}), 
\widehat{P}_{\pm} = \frac{1}{2}( 1 \pm \hat{\gamma}_{5}),
$
and chiral fermion components, 
$\bar{\psi}_{L,R}= \bar{\psi}P_{\pm}, \psi_{R,L}= \widehat{P}_{\pm}\psi$.
The scalar and pseudoscalar densities are given by 
$S(x) = \bar\psi_L \psi_R + \bar\psi_R \psi_L = \bar\psi(1-\frac{a}{2}D)\psi$ and
$P(x) = \bar\psi_L \psi_R - \bar\psi_R \psi_L = \bar\psi\gamma_5(1-\frac{a}{2}D)\psi $.

\FIGURE[h]{
\includegraphics[width=7.5cm]{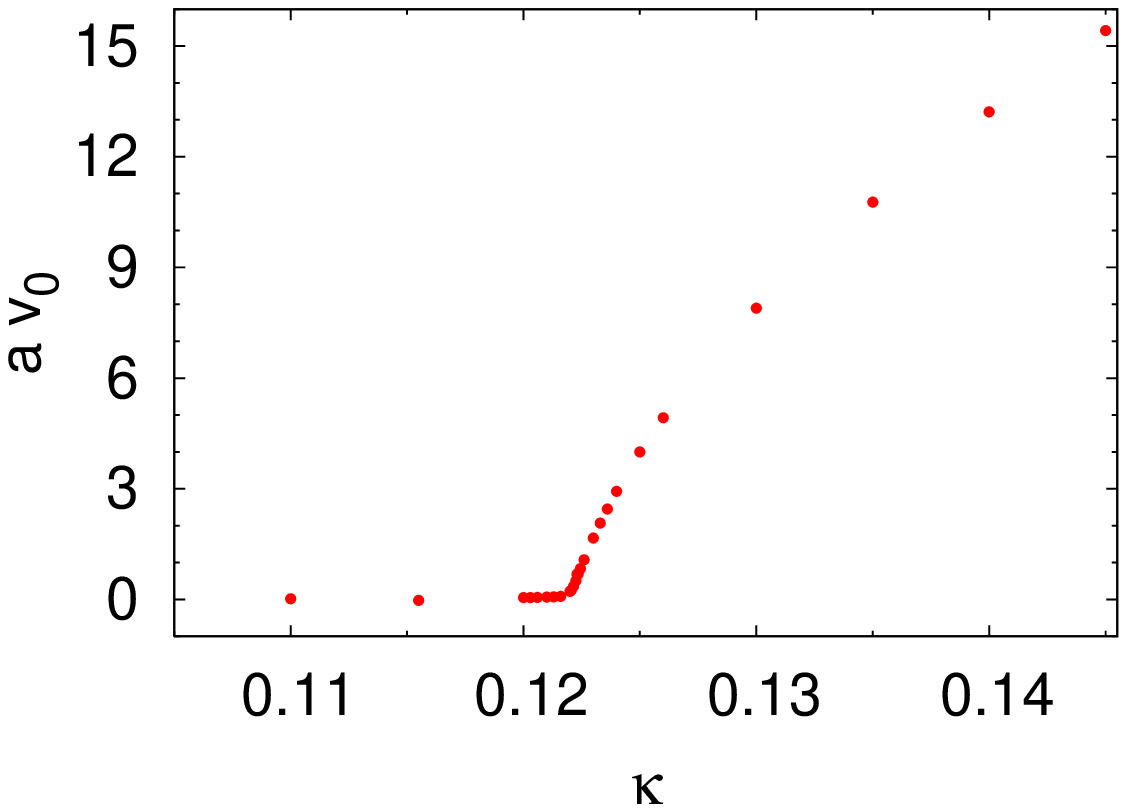}
\caption{\label{fig:GW_phasediagram} 
The vacuum expectation value of the lattice field $\phi_0$
is plotted in lattice spacing units $a$ as a function of the hopping
parameter for fixed values of $\tilde{\lambda}_0 = 10^{-4}$, $\tilde{y}_0=0.35$
with 3 colors of the Top quark. The lattice size is $12^3\times 24$ for the plotted
data. The complete
phase diagram can be mapped out by varying $\tilde{\lambda}_0$ and $\tilde{y}_0$ to determine
$\kappa_c(\tilde{\lambda}_0,\tilde{y}_0)$.
} }

\noindent The Top-Higgs Yukawa model with overlap fermions is defined by the Lagrangian
\begin{eqnarray}
&{\cal L}& = \frac{1}{2} m_0^2 \phi_0^2 + \frac{1}{24} \lambda_0
\phi_0^4 + \frac{1}{2}\left(\partial_\mu \phi_0\right)^2
+ \nn 
&& \bar{\psi}^a_0 \bigl[ D + y_0\cdot\phi_0(1-\frac{a}{2}\cdot D) \bigr]\psi^a_0~, 
\label{eq:L_overlap}
\end{eqnarray}
where the bare overlap fermion field $\psi_0$ and the overlap Dirac operator $D$ were
introduced earlier. Derivatives are represented by finite lattice differences in 
Equation~(\ref{eq:L_overlap}) and
summation over a=1,2,3 for Top color is understood. The gauge link matrices are set to the
unit matrix in Equation~(\ref{eq:D_w}).

The starting point for simulations is
the phase diagram of the theory in the bare coupling space 
of $m_0^2, \lambda_0$, and $y_0$.
The actual location of the critical surface is determined
from the condition $av_0=0$ in a large set
of non-perturbative lattice simulations.
This is shown in Figure~\ref{fig:GW_phasediagram} where the critical
critical hopping parameter for a particular
choice of bare couplings is calculated.
The Higgs part of the lattice Lagrangian is parametrized in the simulations as
$$
{\cal L} = - 2\kappa\sum_\mu \tilde{\phi}_0(x)\tilde{\phi}_0(x+\mu) \\
         + \tilde{\phi}_0^2(x) + \tilde\lambda_0(\tilde{\phi}^2_0(x)-1)^2 ~,
$$
with $\phi_0=\sqrt{2\kappa}\tilde{\phi}_0$, and rescaled notation 
$\tilde{y}_0 = y_0\sqrt{2\kappa}$
for the Yukawa coupling.
The odd number of colors of the single fermion required the application of the 
Rational Hybrid Monte Carlo (RHMC) algorithm for chiral overlap fermion. 
The first new code we developed was based on \cite{Fodor:2003bh,Egri:2006zm}. 
This is the code which
is mostly used in our Top-Higgs-QCD simulations.
We also developed a special FFT version of the RHMC algorithm which exploited the
special structure of the Yukawa coupling in the overlap Dirac operator of the Top-Higgs model.
In the FFT code, Fourier acceleration is used in the evolution of the molecular dynamics 
trajectories which significantly reduced the autocorrelation time 
between independent configurations. 
The details of our RHMC algorithms will be described elsewhere.

\subsection{Comparison of large $\rm{N_F}$ and Monte-Carlo results}

The algorithm was thoroughly tested in the large $N_F$ expansion of the model
where we simulated a sequence of $N_F$ fermions, each with 3 colors, which can also be 
interpreted as the Top quark with $3N_F$  colors. The $N_F\rightarrow\infty$ limit of
the vacuum expectation value $v$ and the Top mass $m_t$ 
were calculated in rescaled $\lambda_0/N_F$ and $y_0/\sqrt{N_F}$ variables for the finite
volumes of the simulations, for fixed value of $m^2_0$. For a particular choice of the rescaled couplings, $v$ and $m_t$
are plotted in Figure~\ref{fig:large_nf_vev} as a function of $1/N_F$. 
\FIGURE[h]{
\includegraphics[width=7.2cm]{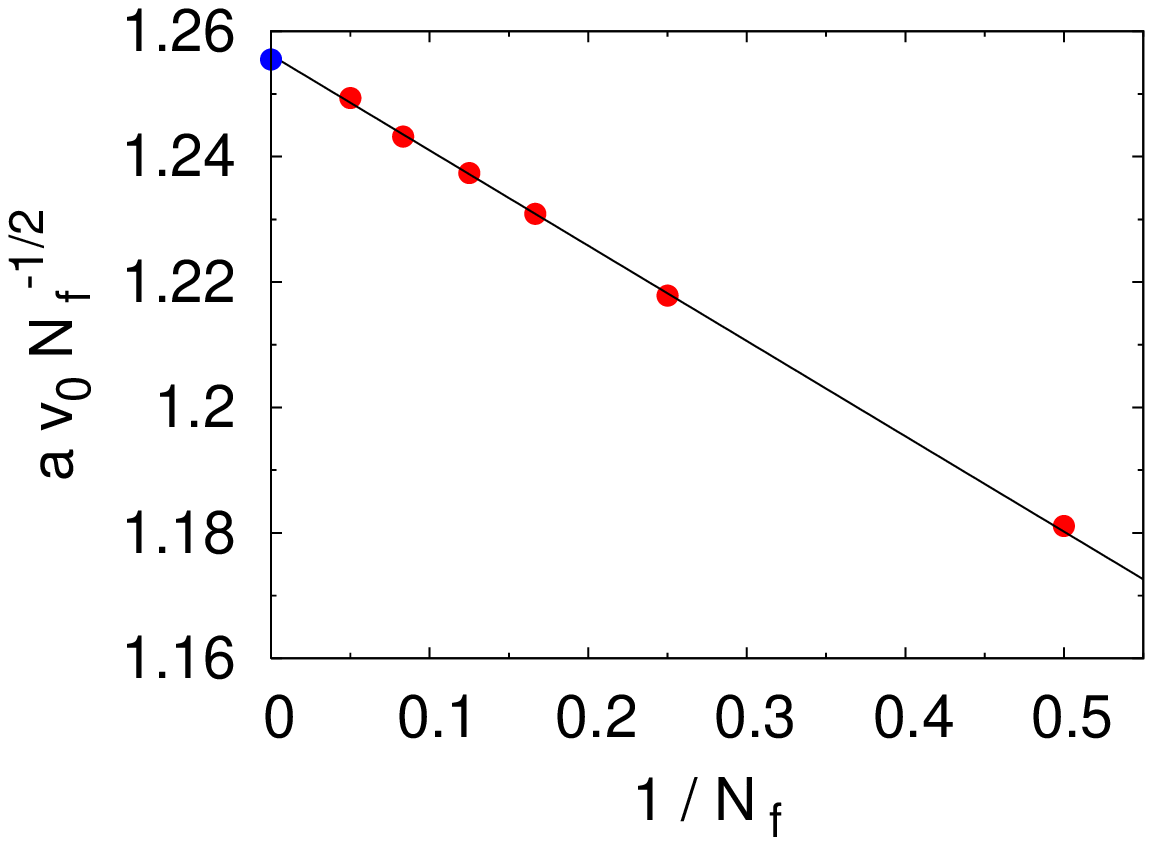}
\includegraphics[width=7.2cm]{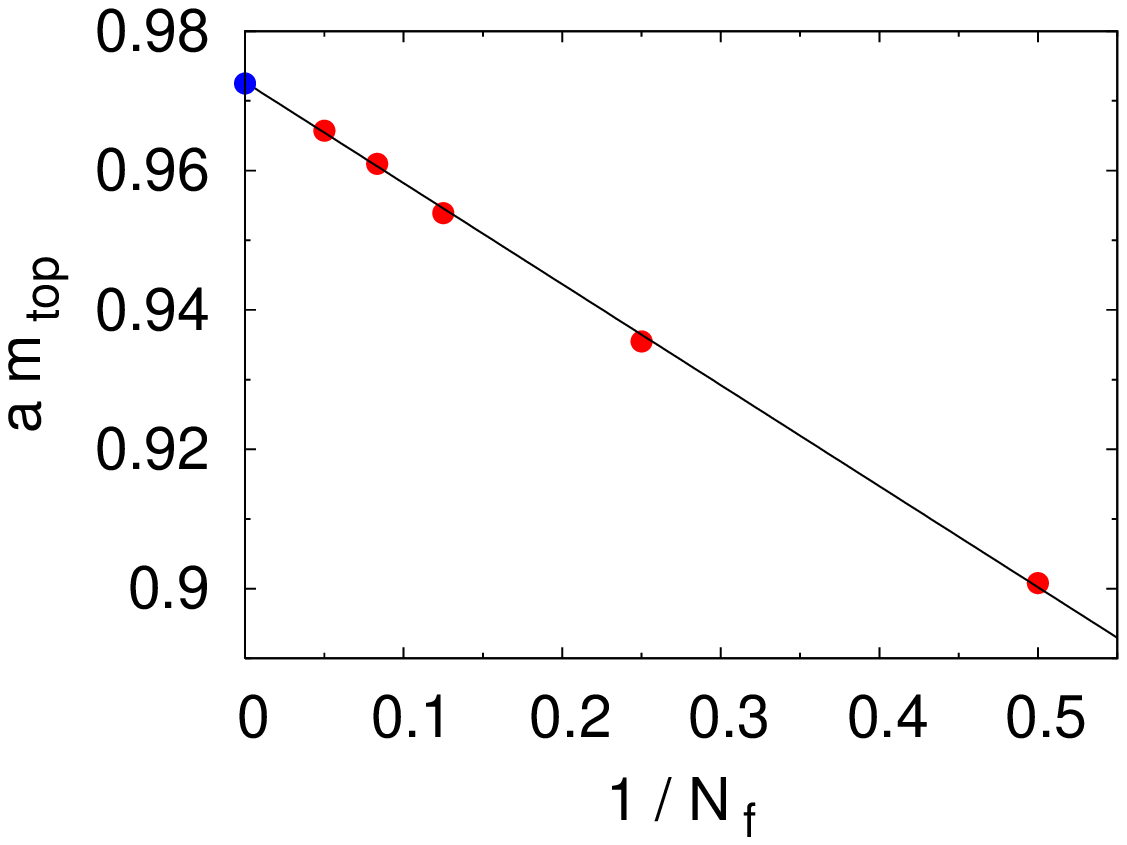}
\caption{\label{fig:large_nf_vev} 
The vacuum expectation value $v$ of the scaled Higgs filed
is plotted on the left as a function of $1/N_F$ for $3N_F$ fermion degrees of freedom.
The blue dot marks the $1/N_F\rightarrow 0$ limit. The right side plot shows the Top
mass as a function of $1/N_F$ with the blue dot marking the 
calculated $1/N_F\rightarrow 0$ limit.
The lattice size was $12^3\times 24$ for every simulation point.
}}
The largest number of
fermions was $3N_F=60$ in the sequence. The solid line indicates the scaled 
asymptotic value of $v$
and $m_t$. The finite $N_F$ data were numerically fitted with an added $1/N_F$ correction term
which allows numerical extrapolation to the $1/N_F\rightarrow 0$ limit with perfect agreement.
For example, in the $vev$ test of Figure~\ref{fig:large_nf_vev} the fitted curve is
$1.2562(4) - 0.152(2)/N_F$ and the large $N_F$ calculation gives $1.2555(7)$ asymptotically,
in excellent agreement with the simulations. The sequence of simulations were done with 
bare parameters $y_0\sqrt{N_F} = 0.7184$, $\lambda_0\cdot N_F = 10^{-3}$, and 
$m^2_0=0.0637$. For the same sequence, the Top quark pole mass $m_t$ was fitted 
on the right side of
Figure~\ref{fig:large_nf_vev} as $0.9727(5) - 0.145(2)/N_F$. The inverse propagator mass
asymptotically is 0.9025 which converts to pole mass $m_t=0.9725$ at the finite
lattice spacing $a$ of the simulations 
by the formula $a m_t= {\rm ln} \frac{2+am}{2-am}$, 
in perfect agreement between simulations and the large $N_F$
prediction.
The complete agreement between the analytic large $N_F$ prediction  
and the Monte-Carlo results provides a very strong cross-check for the 
correctness of our simulation algorithm and the analytic framework.

\subsection{First results on Higgs mass lower bound}

After thorough validation of our algorithm, we turned to a preliminary determination
of the Higgs mass lower bound in the single component Top-Higgs Yukawa model.
The heavy Top quark will constrain the lightest possible
Higgs for any given cutoff in the single component Top-Higgs Yukawa model. 
The starting point for simulations is
the phase diagram of the theory in the bare coupling space 
of $m_0^2, \lambda_0$ and $y_0$.
For every choice of the bare parameter set,
the vacuum expectation value $v$ and the Higgs and Top masses take some values
in lattice cutoff units. Keeping both
the cutoff and the Top mass fixed in physical $vev$ units, we explore all
allowed bare couplings and find the lightest Higgs the theory can
sustain. Repeating this procedure at various distances from the
critical surface determines how the Higgs lower bound varies with the
cutoff.
For the Euclidean path integral to exist, we have to require $\lambda_0 \ge
0$ in the model. We could also consider a
more general Higgs action where the constraint $\lambda_0 \ge
0$ is relaxed when positive terms like $\phi_0^6$ are added in the 
higher-dimensional bare coupling constant space of the bare Lagrangian.
For now we do not include such terms which are part of our ongoing
investigations.

\FIGURE[h]{
\includegraphics[width=7cm]{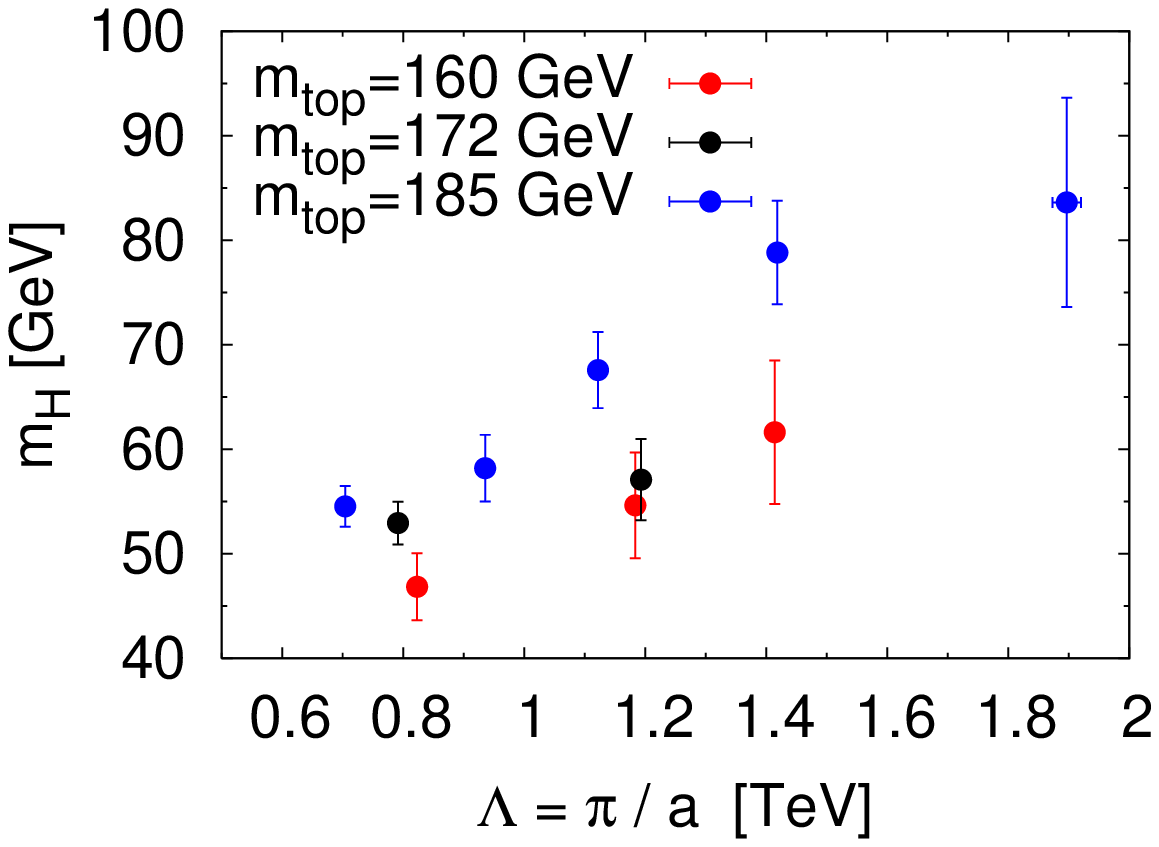}
\caption{\label{fig:overlap_higgs} 
The lowest Higgs mass is plotted as a function of the lattice momentum cutoff for 
three different values of the Top mass. All simulation data are converted to physical units
using $v=246~\rm{GeV}$.
}}
\noindent Figure~\ref{fig:overlap_higgs} displays our preliminary results which 
are not far from what
is expected from the application of the renormalization group. Lattice artifacts 
will require additional interpretation in the low momentum cutoff range of the simulations. 

\noindent {\bf Adding the QCD gauge coupling}

\noindent Our algorithm and simulation code has been extended to the Top-Higgs-QCD code of three
coupling constants. The only change is to include the SU(3) matrix link variables in
the Wilson operator of Equation~(\ref{eq:D_w}) in our construction of the chiral overlap
operator. The numerical determination of the phase diagram and the 
Higgs mass lower bound in the extended  
model with $\lambda,y,g_3$ couplings (Top-Higgs-QCD model) is part of our ongoing Higgs project. 

\section{Higgs mass upper bound and the heavy Higgs particle}

In this section we will review earlier results on the Higgs mass upper bound 
from lattice calculations and illustrate with the higher derivative (Lee-Wick) extension
how a heavy particle might be exhibited without contradictions with Electroweak
precision data.

\subsection{Higgs sector as an effective field theory}

In the Wilsonian view of section 4, the Standard Model is expected to have some yet 
unknown UV completion above a certain energy threshold $\Lambda_0$. 
This threshold could be
as high as $\Lambda_0=M_{\rm{Planck}}$, or as low as $\Lambda_0=1~\rm{TeV}$. 
Below scale $\Lambda_0$ the SM is described 
by the familiar degrees of freedom for the known particles, including fermions and gauge bosons,
in addition to the four-component Higgs field. 
For illustration, we will choose $\Lambda_0=M_{\rm{Planck}}$ first in
the description of the Higgs sector without gauge and Yukawa couplings. 
Generalization to the full 
Standard Model does not add to the purpose of the discussion here. Lowering the cutoff
into the TeV range will be part of the discussion.
If the Higgs sector is treated as an effective theory, the regulator
is chosen for us as an intrinsic part of the theory.
Euclidean four--momenta are smoothly cut off when their lengths
exceed some scale $\Lambda_0$. In this way all momentum
integrals are made manifestly convergent, and no
infinities are encountered. The simplest choice is an exponential 
cutoff function in the propagators,
\be
K_{\Lambda}(p) = \exp \biggl[
-\frac{p^2}{\Lambda_0^2}\biggr]~,
\label{eq:KKK}
\ee
which can be built into the Lagrangian ${\cal L}(\Lambda_0)$ 
for non-perturbative calculations. A mass term could have been added to $p^2$ in
Eq.~(\ref{eq:KKK}) but we simplified the notation for this qualitative discussion.
The general O(4) Higgs Lagrangian at scale 
$\Lambda_0=M_{\rm{Planck}}$ is given by
\bea
{\cal L}_{Higgs} &=& \frac{1}{2} \partial_\mu \phi^a\partial^\mu\phi^a
+ \frac{1}{2}\mu_0^2 \phi^a\phi^a + \frac{\lambda_0}{4!} (\phi^a\phi^a)^2 \nn 
 &+& \frac{c_6}{M_{\rm{Planck}}^2}~\square\phi^a~\square\phi^a
 + \frac{\lambda_6}{M_{\rm{Planck}}^2}(\phi^a\phi^a)^3 +
\frac{c_8}{M_{\rm{Planck}}^4}~\square\partial_\mu\phi^a~\square\partial^\mu\phi^a + ...~,
\label{eq:O(4)_Lagr0}
\eea
where summation is implied over $a=1,2,3,4$.
Only a few higher dimensional operators are included for illustration and
the exponential cutoff is implicitly understood in the functional
integral built on the Lagrangian of Eq.~(\ref{eq:O(4)_Lagr0}).

\subsection{Higgs mass upper bound from the lattice}

The highest allowed Higgs mass from the Lagrangian
of Eq.~(\ref{eq:O(4)_Lagr0}) was investigated before, using lattice cutoff
with $c_6,\lambda_6,c_8$ 
and all other higher dimensional couplings set to zero. Corrections from 
the higher dimensional operators are expected to be small,
of the order of powers of $m_H/M_{\rm{Planck}}$ unless
the couplings $c_6,\lambda_6,c_8$, or any of the other higher dimensional 
couplings are pushed toward asymptotically large values. It is a limit which
is considered artificial and far outside naturalness bounds.  

Convincing evidence
for the Higgs upper bound and its numerical value comes from lattice calculations~\cite{Kuti:1987nr,Luscher:1988gc} where
the derivatives are replaced by finite lattice differences giving up Euclidean
invariance on the Planck scale. The advantage of the lattice approach is that the full $\lambda_0$ range can be scanned from 0 to $\infty$. 
This is important if the Higgs
self-interaction is a marginally (logarithmically) irrelevant operator in the triviality scenario. 
In the limit of infinite cutoff, the largest allowed Higgs mass would be driven to zero (triviality of the renormalized Higgs coupling),
but with the cutoff at the Planck scale we will get a definitive nonvanishing 
upper bound which is saturated at  $\lambda_0=\infty$ in the lattice approximation.
The renormalized Higgs coupling at low energy can be defined as the ratio 
$\lambda_R = 3m_H^2/v^2$ where $v=246~\rm{GeV}$ is the vev of the Higgs field
(the fourth component of the O(4) field), and $m_R$ is a renormalized Higgs propagator
mass which is related in two-loop perturbation theory to the physical Higgs mass
by the relation $m_H=m_R[1+\frac{1}{8192\pi^2}\lambda_R^2]$.
Based on non-perturbative lattice studies, we expect that the largest Higgs
mass is obtained in the $\lambda_0\rightarrow \infty$ limit. For any 
choice of $\lambda_0$ in the O(N) Higgs model we have
\be
m_R=M_{\rm Planck}\cdot C(\lambda_0)\cdot (\beta_1\lambda_R)^{-\frac{\beta_2}{\beta_1}}
{\rm exp}\Big(-\frac{1}{\beta_1\lambda_R}\Big)\Big\{ 1+{\cal O}(\lambda_R) \Big\},
\ee
with $\beta_1=\frac{1}{3}(N+8)\frac{1}{16\pi^2}$ and 
$\beta_2=-\frac{1}{3}(3N+14)\frac{1}{(16\pi^2)^2}$~. The relevant choice is N=4 for the Standard
Model.
The non-universal amplitude $C(\lambda_0)$ is determined from matching to lattice calculations 
in the range $2\pi \le \Lambda/m_H \le 100$~\cite{Kuti:1987nr,Luscher:1988gc}, leading to
the upper bound $m_H=145~\rm{GeV}$ in the $\lambda_0=\infty$ limit, 
if the cutoff is at the Planck scale.
In principle, the lattice cutoff could be replaced by the exponential cutoff function of the 
continuum theory. It would be required to replace the momentum square in Eq.~(\ref{eq:KKK}) by its lattice version and take the inverse lattice spacing much larger than $\Lambda_0$. A new
amplitude would emerge which could change the numerical value of the upper bound without breaking
Euclidean invariance at finite cutoff. This is particularly useful when the cutoff is brought
close to the low energy physical scale. 
In the discussion of
the higher derivative extension of the Higgs sector we will show how to insert a heavy continuum
cutoff scale in the theory which was turned into a practical calculation before~\cite{Jansen:1993jj,Jansen:1993ji}. This suggests that the insertion of the exponential
cutoff scale might be feasible in practical calculations. What remains the most
interesting question for LHC physics is the lowering the cutoff from the Planck scale into
the TeV range. This will be illustrated next in the higher derivative extension of the Higgs
sector with the scale of new physics in the TeV range. 

\subsection{Higher derivative (Lee-Wick) Higgs sector}

An interesting extension of the Standard Model Higgs sector was proposed earlier
by the addition of higher derivative operators using ideas
originally discussed by 
Lee and Wick~\cite{Jansen:1993jj,Jansen:1993ji,Lee:1970iw,Lee:1969fy}. 
Recently a complete Standard Model was constructed
on similar principles~\cite{Grinstein:2007mp}.
Both constructions eliminate
fine tuning in the Higgs sector and require ghost particles on the
TeV scale represented by {\em complex pole pairs} in propagators 
with unusual physical properties. The analysis of the heavy Higgs particle from
~\cite{Jansen:1993jj,Jansen:1993ji} will be followed in our discussion.

In the minimal Standard Model with $SU(2)_L \times U(1)_Y$ gauge
symmetry the Higgs sector is described by a complex
scalar doublet $\Phi$ with quartic self-interaction as we discussed in section 5.
The Higgs potential $V(\Phi^{\dagger}\Phi)$, 
as defined in Equation~\ref{eq:higgs_pot}, is $SU(2)_L \times U(1)_Y$
invariant. It also has a global  $O(4) \approx SU(2)_L \times SU(2)_R$  
symmetry, larger than required by the $SU(2)_L \times U(1)_Y$ gauge symmetry.
Before the weak gauge couplings are switched on, 
it is convenient to represent the Higgs doublet with four real 
components $\phi^a$  which transform in the vector 
representation of $O(4)$. 


We will include new higher derivative terms in the kinetic part of the 
$O(4)$ Higgs Lagrangian,
\begin{equation}
{\cal L}_{H} = \frac{1}{2} \partial_\mu \phi^a\partial^\mu\phi^a
- \frac{{\rm cos}(2\Theta)}{M^2}~\square\phi^a~\square\phi^a
 +
\frac{1}{2M^4}~\square\partial_\mu\phi^a~
\square\partial^\mu\phi^a -
V(\phi^a\phi^a)~,
\label{eq:O(4)_Lagr1}
\end{equation}
where summation is implied over $a=1,2,3,4$. 
Also, in this subsection and the next, we use the Minkowski metric
and a familiar, convenient form of the Higgs potential,
$
V(\phi^a\phi^a) = - \frac{1}{2}\mu^2 \phi^a\phi^a +
\lambda (\phi^a\phi^a)^2.
$
The higher derivative terms of the Lagrangian in
Eq.~(\ref{eq:O(4)_Lagr1}) lead to complex conjugate ghost pairs in the 
spectrum of the Hamilton operator. The complex conjugate pairs of energy
eigenvalues of the Hamilton operator and the related complex pole pairs in the
propagator of the scalar field ghost particles are
parametrized by ${\cal M}=M e^{{\pm}i\Theta}$. The absolute value $M$
of the complex ghost mass  ${\cal M}$ will be set on the TeV scale. 
The Higgs Lagrangian ${\cal L}_H$ in Equation~(\ref{eq:O(4)_Lagr1}) describes a
finite field theory without divergences, or fine tuning. 
It has a particularly simple form with the special choice $\Theta = \pi/4$
of the complex ghost phase, 
\begin{equation}
{\cal L}_{H} = \frac{1}{2} \partial_\mu \phi^a\partial^\mu\phi^a
 + \frac{1}{2M^4}~\square\partial_\mu\phi^a~\square\partial^\mu\phi^a -
 V(\phi^a\phi^a)~.
\label{eq:O(4)_Lagr2}
\end{equation}
The $\Theta \rightarrow 0$ limit in Eq.~(\ref{eq:O(4)_Lagr1}) requires 
special attention. In this limit, 
the ghost particle becomes real and to avoid a double real pole in the
propagator with problematic behavior, the choice $\Theta = 0$ 
requires to drop the 
$
\frac{1}{2M^4}~\square\partial_\mu\phi^a~\square\partial^\mu\phi^a
$ 
derivative term in the Lagrangian,
\begin{equation}
{\cal L}_{H} = \frac{1}{2} \partial_\mu \phi^a\partial^\mu\phi^a
 - \frac{1}{2M^2}~\square\phi^a~\square\phi^a -
 V(\phi^a\phi^a)~,
\label{eq:O(4)_Lagr3}
\end{equation}
the starting point of~\cite{Grinstein:2007mp}. 
\subsection{Gauge and Yukawa couplings}
Gauging the Lagrangian~(\ref{eq:O(4)_Lagr2}) remained unpublished before~\cite{Kuti:1}. For completeness, we present the main results.
The construction of the higher derivative U(1) gauge Lagrangian mirrors
Eq.~(\ref{eq:O(4)_Lagr2}) for the special choice $\Theta=\pi/4$, 
\begin{equation}
{\cal L_{{\rm B}}} = - \frac{1}{4}F_{\mu\nu} F^{\mu\nu}
- \frac{1}{4M^4} \Box F_{\mu\nu} \Box F^{\mu\nu}~,
\label{eq:Lag_u1}
\end{equation}
with U(1) gauge field $B_\mu$ and
$F_{\mu\nu}=\partial_\mu B_\nu - \partial_\nu B_\mu$.
In addition to the massless gauge vector boson, the higher derivative term in 
Eq.~(\ref{eq:Lag_u1}) will insert a ghost particle in the spectrum
of the Hamiltonian with a complex conjugate pole pair parametrized 
by ${\cal M}=Me^{\pm i\Theta}$. For a general complex phase $\Theta$ an
additional term will appear in the Lagrangian, 
in close analogy with the construction of Eq.~(\ref{eq:O(4)_Lagr1}).

The higher derivative Yang-Mills gauge Lagrangian for the $SU(2)_W$ 
weak gauge field $W_\mu$ will follow a similar construction adding the 
dimension eight ghost term, 
\begin{equation}
{\cal L_{{\rm W}}} = - \frac{1}{4}G_{\mu\nu}^a G^{a\mu\nu}
- \frac{1}{4M^4}D^2G_{\mu\nu}^a D^2 G^{a\mu\nu}~,
\label{eq:Lag_su2}
\end{equation}
where the notation $G_{\mu\nu}^a=\partial_\mu W^a_\nu - \partial_\nu W^a_\mu +
g f^{abc} W_\mu^b W_\nu^c$ is used with the covariant derivative 
$D_\mu^{ab} = \delta^{ab}\partial_\mu + g f^{abc} W_\mu^c$.
Higher derivative Lagrangians, similar to Eq.~(\ref{eq:Lag_su2}), were first
introduced by Slavnov to regulate Yang-Mills theories~\cite{Bakeyev:1996is}.

Labeling the components of the complex $SU(2)_L$ Higgs-doublet field
as 
$
\Phi=\binom{\Phi^+}{\Phi^0}
$ 
the gauged Higgs sector is described by the Lagrangian
${\cal L} = {\cal L}_{W} +{\cal L}_{B} + {\cal L}_{Higgs}$ with the Higgs
Lagrangian 
\begin{equation}
{\cal L}_{Higgs}=(D_\mu\Phi)^\dagger D^\mu\Phi  +
\frac{1}{2M^4}(D_\mu D^\dagger D\Phi)^\dagger (D_\mu D^\dagger D\Phi) -
V(\Phi^\dagger\Phi)
\label{eq:Higgs}
\end{equation}
where the Higgs potential is
$ V(\Phi^\dagger\Phi)=-\frac{1}{2}\mu^2\Phi^\dagger\Phi+\lambda
(\Phi^\dagger\Phi)^2$ and the gauge-covariant derivative is
$D_\mu\Phi=\left(\partial_\mu+i\frac{g}{2}\sigma\cdot
W_\mu+i\frac{g'}{2}B_\mu\right)\Phi$.
The higher derivative term in the fermion Lagrangian will take the form
\begin{equation}
{\cal L}_{fermion} = i\overline{\Psi} {\Dslash} ~ 
\Psi +\frac{i}{2M^4} \overline{\Psi}~\Dslash^2 \Dslash \Dslash^2 ~ \Psi.
\label{eq:Dirac}
\end{equation}

Next we will briefly summarize two important features of the higher derivative Higgs sector
with the ghost mass scale in the TeV range. The RG running of the Higgs coupling 
freezes asymptotically
and a much heavier Higgs particle is allowed in extended Higgs dynamics.

\subsection{Running Higgs coupling in the higher derivative Higgs sector}
\FIGURE[h]{
\includegraphics[width=6cm]{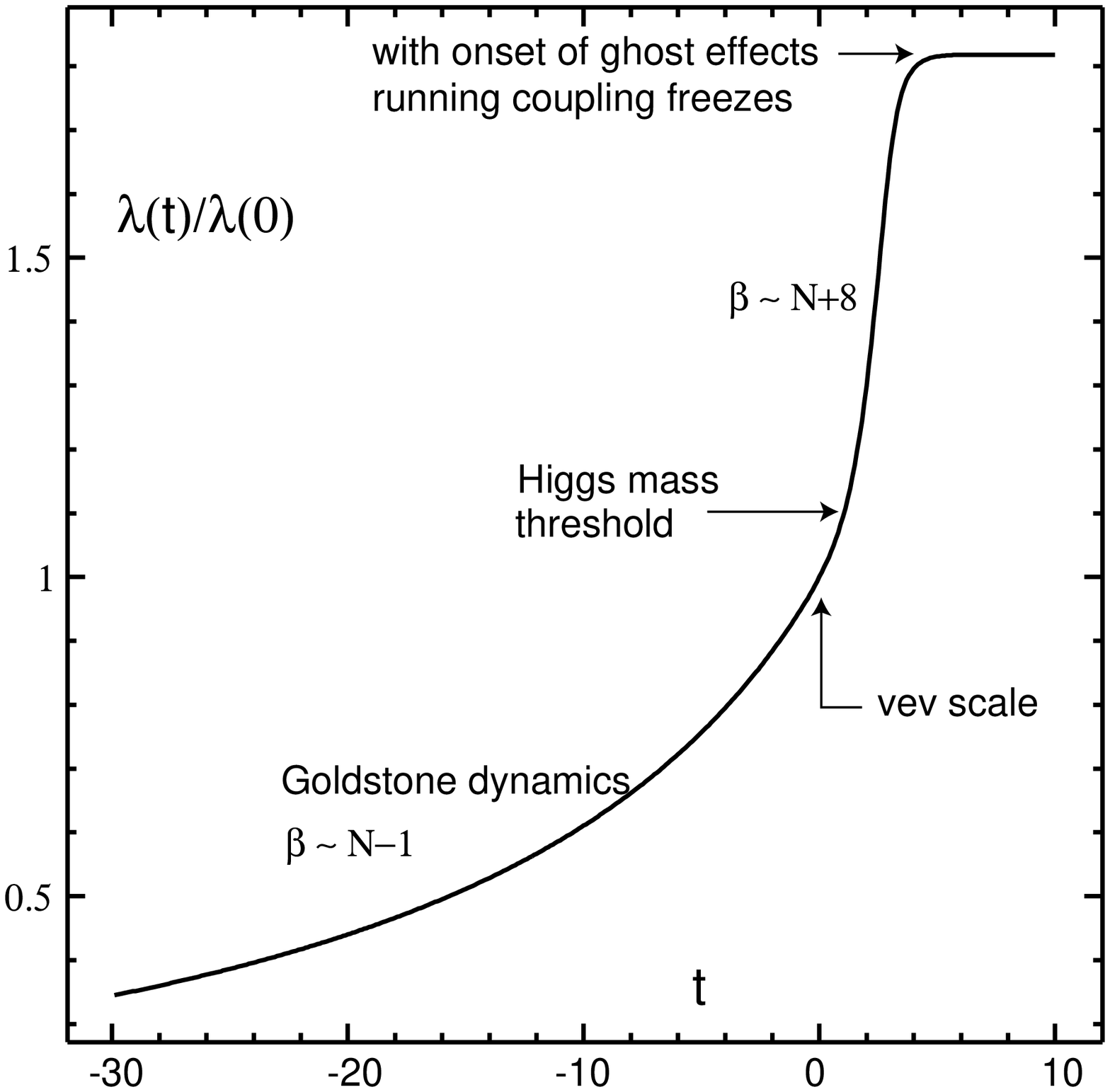}
\caption{\label{fig:running_lambda} Running Higgs coupling in the higher \hyphenation{deriv-ative} Higgs sector. 
}}
This can be illustrated by calculating the scale dependent 
one-loop $\beta$-function within renormalized perturbation theory in the 
broken phase of the higher derivative O(N) Higgs sector~\cite{Liu:1994zy,Kuti:1994ii}.
In addition to N-1 massless Goldstone modes, there is a massive Higgs 
excitation and a massive 
complex conjugate ghost pair appears in all N channels, as a consequence of the 
new derivative term in the Lagrangian.
On a low energy scale $\mu$, when $t={\rm log}(\mu/v)$ is negative, 
the $\beta$-function is dominated by the Goldstone modes whose one-loop
contribution is $\frac{N-1}{2\pi^2}\lambda^2(t)$. Above the Higgs 
mass threshold the massive Higgs loop contribution sets in
and the $\beta$-function becomes $\frac{N+8}{2\pi^2}\lambda^2(t)$ which is
the familiar one-loop form in the minimal mass 
independent subtraction scheme of the standard O(N) model.
As $t$ increases, the complex ghost loop becomes
increasingly important and well beyond
the ghost scale M, for $t\gg {\rm log}(M/vev)$, the beta-function will asymptotically vanish. 
The running coupling constant $\lambda(t)$ 
first will grow as $t$ increases, but eventually it will freeze at some 
asymptotic value $\lambda(\infty)$ as shown in Fig.~\ref{fig:running_lambda}.
Ghost loops in the higher derivative Higgs model cancel the loops effects
from the low-energy SM particles in the UV region and this `anti-screening'
effect opens up the possibility for such theories to be more strongly 
interacting than the standard Higgs sector.

\subsection{Scattering amplitudes}

The Higgs particle is defined as the resonance pole in the s-channel 
Goldstone scattering amplitude. The Goldstone amplitude
can be calculated in the higher derivative
Higgs sector of the O(N) Lagrangian in the large N approximation. In addition, the Higgs
particle can be investigated directly in lattice simulations of 
the higher derivative model, just like in the
standard Higgs sector. 

\begin{figure}[h]
\vskip 0.5cm
\centerline{ \epsfysize=4.0cm 
             \epsfxsize=4.0cm \epsfbox{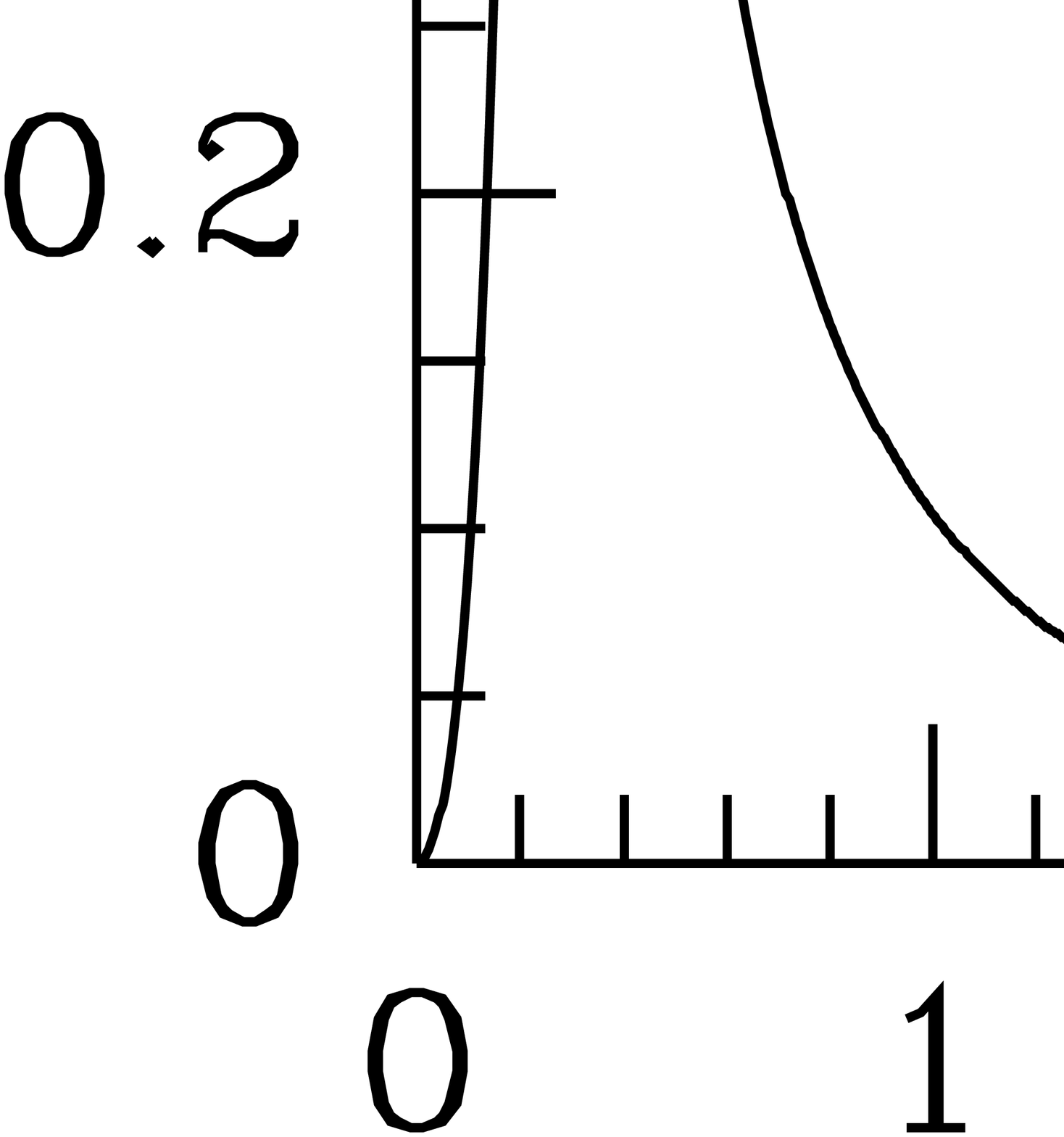}}
\vskip 1cm
\caption{ The Goldstone Goldstone scattering cross section and phase
shift is plotted against the center of mass energy in large-$N$
expansion for the Pauli-Villars higher derivative
$O(N)$ theory. The input vev value is $v=0.07$ in $M$ units.
 The peak corresponds to the Higgs resonance, which is at
$m_H=0.28$ in $M$ units. The scattering cross section is completely
smooth across the so-called ghost pole locations. }
\label{fig:ch3.scatter}
\end{figure}

In Figure~\ref{fig:ch3.scatter} we plotted from ~\cite{Liu:1994zy,Liu:1994uh} 
the cross section
as a function of the $\sqrt{s}$ center of mass energy in ghost mass units. 
The location of the complex
Goldstone ghost pair in the scattering amplitude of the first Riemann sheet
is determined by the choice of the phase angle $\Theta=\pi/4$ 
in the Lagrangian of Equation~(\ref{eq:O(4)_Lagr1}). The peak in the cross section
corresponds to the complex Higgs resonance pole on the second sheet of the
scattering amplitude. 
%

Also plotted in Figure~\ref{fig:ch3.scatter}
is the scattering phase shift as a function of $\sqrt{s}$. The phase shift 
has a sharp rise at the Higgs pole; however the cross section and the shape of
the phase shift do not describe a standard Breit-Wigner shape in the presence
of the ghosts and higher derivative Higgs dynamics.  
It is `unusual' that the phase shift decreases as the
energy gets through the real part of the ghost mass signaling 
acausal behavior in the scattering amplitude.
It had been argued by Lee that this acausal
behavior would only  occur on microscopic scales,
typical of the Compton wave length of ghosts, and it will not lead to
macroscopic acausal observations.
In the large N plots of Figure~\ref{fig:ch3.scatter}, the bare parameters were 
tuned to $m_H= 1~\rm{TeV}$ for the Higgs mass with the ghost threshold located at 
3.6 TeV. Lattice simulations confirmed similar strongly interacting 
heavy Higgs physics scenarios~\cite{Liu:1994zy,Kuti:1994ii}.

\subsection{Heavy Higgs particle and the $\rho$-parameter}

We discussed in the introduction that a heavy Higgs particle, beyond the 200 GeV range, is not consistent with Electroweak precision data in the perturbative sense. Concerns were raised earlier
that the heavy Higgs particle of the higher derivative Higgs sector will contribute to the
Electroweak $\rho$-parameter beyond experimentally allowed limits~\cite{Chivukula:1996sn}.
Straightforward application
\FIGURE[h]{
\includegraphics[width=8cm]{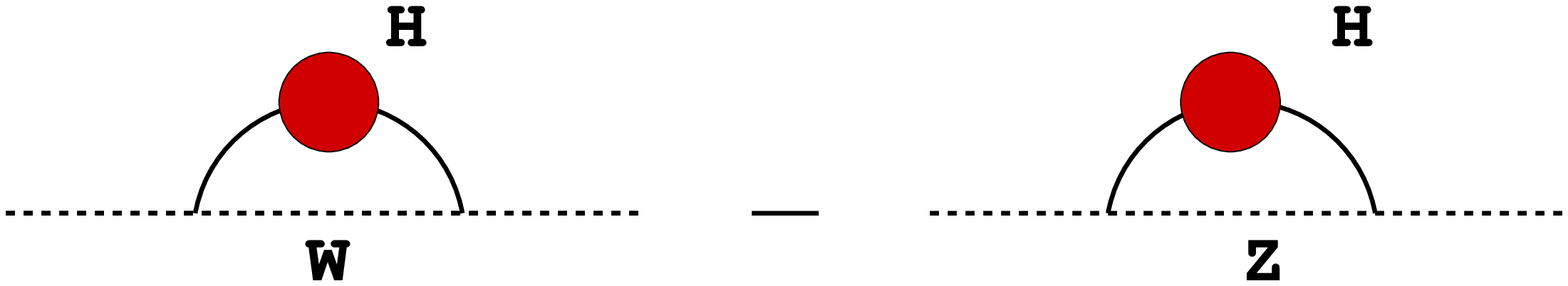}
\caption{\label{fig:rho_parameter} Higgs contribution to electroweak 
vacuum polarization operator.
}}
\noindent of perturbative loop integrals support this concern.
However, with new physics on the TeV scale (represented by ghost particles) the loop integrals
are considerably different. A crude estimate can be made by evaluating 
the contribution of the vacuum polarization tensors $\Pi^H_W,\Pi^H_Z$ to the $\rho$-parameter, 
\begin{equation*}
\rho -1 |_{\rm Higgs} = \frac{\Pi^H_W}{M^2_{W,tree}}-\frac{\Pi^H_Z}{M^2_{Z,tree}}
                      = - \frac{3}{4}g'^{2}\int_{k^2<\Lambda^2} \frac{d^4k}{(2\pi)^4}
\frac{\Sigma_H(k^2)}{(k^2 + M^2_{W,tree})(k^2 + M^2_{Z,tree})(k^2 + \Sigma_H(k^2))}~,
\end{equation*}
with a sharp momentum cutoff in the TeV range and using the tree level Higgs self-energy
operator $\Sigma_H(k^2) $. The reduction is quite large in comparison with the 1-loop
perturbative formula. Replacing the cutoff integral by the Pauli-Villars regulator, which
is appropriate for the higher derivative theory, we get similar reduction. The effects of
the non-perturbative Higgs dynamics represented by a complicated  $\Sigma_H(k^2) $ operator
would have to be determined by non-perturbative simulations. If these reduction effects are not
sufficient, one might need to add another Higgs doublet to the extended Higgs
sector in the spirit of recent suggestions~\cite{Barbieri:2006dq}.
To exhibit a heavy Higgs particle as a broad resonance, with strong interaction and with 
acceptable $\rho$-parameter, remains an interesting challenge for lattice Higgs physics
and model building.

\subsection*{Acknowledgements}
J.K. is greatful for interesting discussions with J.~Espinosa and
D.~N. would like to 
acknowledge helpful discussions with C.~Hoelbling and K.~Szabo.
This research was supported by the DOE under grants DOE-FG03-97ER40546,
DE-FG02-97ER25308, by the NSF under grant 0704171, by DFG under grant FO 502/1, 
and by the EU under grant I3HP.


\begin{thebibliography}{99}
\bibitem{Barate:2003sz}
  R.~Barate {\it et al.}  [LEP Working Group for Higgs boson searches],
  Phys.\ Lett.\ B {\bf 565}, 61 (2003),
  {\tt arXiv:hep-ex/0306033}.

\bibitem{LEP2005}
 LEP Electroweak Working Group, {\tt http://lepewwg.web.cern.ch/LEPEWWG/}.
    
\bibitem{Chanowitz:2002cd}
  M.~S.~Chanowitz,
  Phys.\ Rev.\ D {\bf 66}, 073002 (2002),
  {\tt arXiv:hep-ph/0207123}.

\bibitem{Peskin:2001rw}
  M.~E.~Peskin and J.~D.~Wells,
  Phys.\ Rev.\  D {\bf 64}, 093003 (2001),
  {\tt arXiv:hep-ph/0101342}.

\bibitem{Grojean:2006nn}
  C.~Grojean, W.~Skiba and J.~Terning,
  Phys.\ Rev.\  D {\bf 73}, 075008 (2006),
  {\tt arXiv:hep-ph/0602154}.

\bibitem{Hagiwara:2002fs}
  K.~Hagiwara {\it et al.}  [Particle Data Group],
  Phys.\ Rev.\ D {\bf 66}, 010001 (2002).

\bibitem{Hambye:1996wb}
  T.~Hambye and K.~Riesselmann,
  Phys.\ Rev.\ D {\bf 55}, 7255 (1997), 
  {\tt arXiv:hep-ph/9610272}.

\bibitem{Casas:1996aq}
  J.~A.~Casas, J.~R.~Espinosa and M.~Quiros,
  Phys.\ Lett.\ B {\bf 382}, 374 (1996).

\bibitem{Holland:2003jr}
  K.~Holland and J.~Kuti,
  Nucl.\ Phys.\ Proc.\ Suppl.\  {\bf 129}, 765 (2004),
  {\tt arXiv:hep-lat/0308020}.

\bibitem{Holland:2004sd}
  K.~Holland,
  Nucl.\ Phys.\ Proc.\ Suppl.\  {\bf 140}, 155 (2005),
  {\tt arXiv:hep-lat/0409112}.

\bibitem{Jansen:1993jj}
  K.~Jansen, J.~Kuti and C.~Liu,
  Phys.\ Lett.\  B {\bf 309}, 119 (1993),
  {\tt arXiv:hep-lat/9305003}.

\bibitem{Jansen:1993ji}
  K.~Jansen, J.~Kuti and C.~Liu,
  Phys.\ Lett.\  B {\bf 309}, 127 (1993),
  {\tt arXiv:hep-lat/9305004}.

\bibitem{Grinstein:2007mp}
  B.~Grinstein, D.~O'Connell and M.~B.~Wise,
  {\tt arXiv:0704.1845 [hep-ph]}.

\bibitem{Coleman:1973jx}
  S.~R.~Coleman and E.~Weinberg,
  Phys.\ Rev.\ D {\bf 7}, 1888 (1973).

\bibitem{Kuti:1987bs}
  J.~Kuti and Y.~Shen,
  Phys.\ Rev.\ Lett.\  {\bf 60}, 85 (1988).

\bibitem{Fukuda:1974ey}
  R.~Fukuda and E.~Kyriakopoulos,
  Nucl.\ Phys.\ B {\bf 85}, 354 (1975).

\bibitem{O'Raifeartaigh:1986hi}
  L.~O'Raifeartaigh, A.~Wipf and H.~Yoneyama,
  Nucl.\ Phys.\ B {\bf 271}, 653 (1986).

\bibitem{Susskind:1976jm}
  L.~Susskind,
  Phys.\ Rev.\ D {\bf 16}, 3031 (1977).

\bibitem{Kogut:1974ag}
  J.~B.~Kogut and L.~Susskind,
  Phys.\ Rev.\ D {\bf 11}, 395 (1975).

\bibitem{Duane:1987de}
  S.~Duane, A.~D.~Kennedy, B.~J.~Pendleton and D.~Roweth,
  Phys.\ Lett.\ B {\bf 195}, 216 (1987).

\bibitem{Wilson:1973jj}
  K.~G.~Wilson and J.~B.~Kogut,
  Phys.\ Rept.\  {\bf 12}, 75 (1974).

\bibitem{Krive:1976sg}
  I.~V.~Krive and A.~D.~Linde,
  Nucl.\ Phys.\ B {\bf 117}, 265 (1976).

\bibitem{Linde:1977mm}
  A.~D.~Linde,
  Phys.\ Lett.\ B {\bf 70}, 306 (1977).

\bibitem{Politzer:1978ic}
  H.~D.~Politzer and S.~Wolfram,
  Phys.\ Lett.\ B {\bf 82}, 242 (1979)
  [Erratum-ibid.\  {\bf 83B}, 421 (1979)].

\bibitem{Cabibbo:1979ay}
  N.~Cabibbo, L.~Maiani, G.~Parisi and R.~Petronzio,
  Nucl.\ Phys.\ B {\bf 158}, 295 (1979).

\bibitem{Hung:1979dn}
  P.~Q.~Hung,
  Phys.\ Rev.\ Lett.\  {\bf 42}, 873 (1979).

\bibitem{Flores:1982rv}
  R.~A.~Flores and M.~Sher,
  Phys.\ Rev.\ D {\bf 27}, 1679 (1983).

\bibitem{Lindner:1985uk}
  M.~Lindner,
  Z.\ Phys.\ C {\bf 31}, 295 (1986).

\bibitem{Sher:1988mj}
  M.~Sher,
  Phys.\ Rept.\  {\bf 179}, 273 (1989).

\bibitem{Lindner:1988ww}
  M.~Lindner, M.~Sher and H.~W.~Zaglauer,
  Phys.\ Lett.\ B {\bf 228}, 139 (1989).

\bibitem{Ford:1992mv}
  C.~Ford, D.~R.~T.~Jones, P.~W.~Stephenson and M.~B.~Einhorn,
  Nucl.\ Phys.\ B {\bf 395}, 17 (1993).

\bibitem{Sher:1993mf}
  M.~Sher,
  Phys.\ Lett.\ B {\bf 317}, 159 (1993)
  [Addendum-ibid.\ B {\bf 331}, 448 (1994)].

\bibitem{Casas:1994qy}
  J.~A.~Casas, J.~R.~Espinosa and M.~Quiros,
  Phys.\ Lett.\ B {\bf 342}, 171 (1995).

\bibitem{Altarelli:1994rb}
  G.~Altarelli and G.~Isidori,
  Phys.\ Lett.\ B {\bf 337}, 141 (1994).

\bibitem{Boyanovsky:1997dj}
  D.~Boyanovsky, W.~Loinaz and R.~S.~Willey,
  Phys.\ Rev.\ D {\bf 57}, 100 (1998).

\bibitem{Einhorn:2007rv}
  M.~B.~Einhorn and D.~R.~T.~Jones,
  JHEP {\bf 0704}, 051 (2007),
  {\tt arXiv:hep-ph/0702295}.

\bibitem{Arnold:1989cb}
  P.~B.~Arnold,
  Phys.\ Rev.\ D {\bf 40}, 613 (1989).

\bibitem{Anderson:1990aa}
  G.~W.~Anderson,
  Phys.\ Lett.\ B {\bf 243}, 265 (1990).

\bibitem{Arnold:1991cv}
  P.~Arnold and S.~Vokos,
  Phys.\ Rev.\ D {\bf 44}, 3620 (1991).

\bibitem{Espinosa:1995se}
  J.~R.~Espinosa and M.~Quiros,
  Phys.\ Lett.\ B {\bf 353}, 257 (1995),
  {\tt arXiv:hep-ph/9504241}.

\bibitem{Isidori:2001bm}
  G.~Isidori, G.~Ridolfi and A.~Strumia,
  Nucl.\ Phys.\ B {\bf 609}, 387 (2001),
  {\tt arXiv:hep-ph/0104016}.

\bibitem{Gerhold:2007gx}
  P.~Gerhold and K.~Jansen,
  {\tt arXiv:0707.3849 [hep-lat]}.

\bibitem{Gerhold:2007yb}
  P.~Gerhold and K.~Jansen,
  JHEP {\bf 0709}, 041 (2007),
  {\tt arXiv:0705.2539 [hep-lat]}.

\bibitem{Lin:1990ue}
  L.~Lin, I.~Montvay, H.~Wittig and G.~Munster,
  Nucl.\ Phys.\ B {\bf 355}, 511 (1991).

\bibitem{Bock:1990tv}
  W.~Bock, A.~K.~De, K.~Jansen, J.~Jersak, T.~Neuhaus and J.~Smit,
  Nucl.\ Phys.\ B {\bf 344}, 207 (1990).

\bibitem{Lee:1989mi}
  I.~H.~Lee, J.~Shigemitsu and R.~E.~Shrock,
  Nucl.\ Phys.\ B {\bf 334}, 265 (1990).

\bibitem{Fodor:2003bh}
  Z.~Fodor, S.~D.~Katz and K.~K.~Szabo,
  JHEP {\bf 0408}, 003 (2004),
  {\tt arXiv:hep-lat/0311010}.

\bibitem{Egri:2006zm}
G.~I.~Egri, Z.~Fodor, C.~Hoelbling, S.~D.~Katz, D.~Nogradi and K.~K.~Szabo,
Comput.\ Phys.\ Commun.\  {\bf 177} (2007) 631,
{\tt arXiv:hep-lat/0611022}.


\bibitem{Kuti:1987nr}
  J.~Kuti, L.~Lin and Y.~Shen,
  Phys.\ Rev.\ Lett.\  {\bf 61}, 678 (1988).

\bibitem{Luscher:1988gc}
  M.~Luscher and P.~Weisz,
  Phys.\ Lett.\  B {\bf 212}, 472 (1988).

\bibitem{Lee:1970iw}
  T.~D.~Lee and G.~C.~Wick,
  Phys.\ Rev.\  D {\bf 2}, 1033 (1970).

\bibitem{Lee:1969fy}
  T.~D.~Lee and G.~C.~Wick,
  Nucl.\ Phys.\  B {\bf 9}, 209 (1969).

\bibitem{Kuti:1}
J.~Kuti,
{\em Unpublished.}

\bibitem{Bakeyev:1996is}
  T.~D.~Bakeyev and A.~A.~Slavnov,
  Mod.\ Phys.\ Lett.\  A {\bf 11}, 1539 (1996),
  {\tt arXiv:hep-th/9601092}
{\em (this paper contains relevant earlier references).}

\bibitem{Liu:1994zy}
  C.~Liu,
  {\tt arXiv:0704.3999 [hep-ph]}.

\bibitem{Kuti:1994ii}
  J.~Kuti,
  Nucl.\ Phys.\ Proc.\ Suppl.\  {\bf 42}, 113 (1995),
  {\tt arXiv:hep-lat/9502018}.


\bibitem{Liu:1994uh}
  C.~Liu, K.~Jansen and J.~Kuti,
  Nucl.\ Phys.\ Proc.\ Suppl.\  {\bf 42}, 630 (1995),
  {\tt arXiv:hep-lat/9412034}.

\bibitem{Chivukula:1996sn}
  R.~S.~Chivukula and E.~H.~Simmons,
  Phys.\ Lett.\  B {\bf 388}, 788 (1996),
  {\tt arXiv:hep-ph/9608320}.

\bibitem{Barbieri:2006dq}
  R.~Barbieri, L.~J.~Hall and V.~S.~Rychkov,
  Phys.\ Rev.\  D {\bf 74}, 015007 (2006),
  {\tt arXiv:hep-ph/0603188}.



\end{thebibliography}
\end{document}